\documentclass[prb,twocolumn,showpacs,floatfix]{revtex4}
\usepackage{graphicx}
\usepackage{amssymb}
\usepackage{amsmath}
\usepackage{xspace}
\usepackage{url}

\begin{document} 
\title{Existence of the excitonic insulator phase 
in the extended Falicov-Kimball model: 
an $SO(2)$-invariant slave-boson approach}
\author{B. Zenker$^{1}$,   D. Ihle$^{2}$, F. X. Bronold$^{1}$, and 
H. Fehske$^{1}$}
\affiliation{$^{1}$Institut f{\"ur} Physik,
             Ernst-Moritz-Arndt-Universit{\"a}t Greifswald,
             17489 Greifswald,
             Germany\\$^{2}$Institut f\"ur Theoretische Physik, 
Universit\"at Leipzig, 04109 Leipzig, Germany}




\date{\today}
\begin{abstract}
We re-examine the three-dimensional spinless Falicov-Kimball model 
with dispersive $f$ electrons at half-filling, addressing the dispute about 
the formation of an excitonic condensate, which is closely related to the 
problem of electronic ferroelectricity.  
To this end, we work out a slave-boson functional integral representation 
of the suchlike extended Falicov-Kimball model that preserves the 
$SO(2)\otimes U(1)^{\otimes 2}$  invariance of the action.
We find  a spontaneous pairing of $c$ electrons with $f$ holes, 
building an excitonic insulator state 
at low temperatures, also for the case of initially non-degenerate 
orbitals. This is in contrast to recent predictions of scalar 
slave-boson mean-field theory
but corroborates previous Hartree-Fock and RPA results.
Our more precise treatment of correlation effects, however, leads 
to a substantial reduction 
of the critical temperature.  
The different behavior of the partial densities of 
states in the  weak and strong 
inter-orbital Coulomb interaction regimes supports 
a BCS-BEC transition scenario. 
\end{abstract}
\pacs{71.28.+d, 71.35.-y, 71.35.Lk, 71.30.+h, 71.28.+d. 71.27.+a}
\maketitle
\section{Introduction}
The excitonic instability in solids is driven by the Coulomb attraction 
between electrons and holes which under certain conditions causes 
them to form bound states. At the semimetal-semiconductor
transition the conventional ground state of the crystal may become unstable
with respect to the spontaneous formation of excitons. 
Starting from a semimetal with a sufficiently small number of electrons 
and holes, such that the Coulomb interaction is basically unscreened, the 
number of free carriers will vary discontinuously under an applied 
perturbation~\cite{Mo61}, signaling a phase transition. 
Approaching the transition from 
the semiconductor side, an anomaly occurs when the (indirect) 
band gap, tuned, e.g., by external pressure, becomes less then the 
exciton binding energy~\cite{Kno63}. As a consequence, a new distorted
phase of the crystal, with spontaneous coherence between conduction
and valence bands and a gap for charged excitations, develops. It  
separates, below a critical temperature, the semimetal from the 
semiconductor. This state, known as `excitonic insulator' (EI),
can be regarded as an electron-hole pair (exciton) condensate~\cite{HR68}.  
By nature, depending on from which side of the semimetal-semiconductor 
transition the EI is approached, the EI typifies either as a BCS condensate 
of loosely bound electron-hole pairs or as a Bose-Einstein  
condensate (BEC) of preformed tightly-bound excitons~\cite{Le80,CN82,BF06}.

The mean-field description of the EI phase is very similar to 
the BCS theory of superconductivity, and has been worked out long-time 
ago~\cite{Cl65,Ko68,KK65,KM65,JRK67,Zi68}. In this context also
the transition from BCS to BE condensation was 
discussed~\cite{Le80,NS85,ZS85,KSH08}. Surprisingly enough, the quantitative
semimetal-EI-semiconductor phase diagram has been determined only quite 
recently~\cite{BRF07,BF08,KSH08}. All these investigations,
having normal (excited) semiconductor systems in mind, rest on the standard 
effective-mass Mott-Wannier-type exciton model. Thereby important 
band-structure and correlation effects, as well as the exciton-excition
and exciton-phonon interaction, and the inter-valley scattering of excitons 
were largely neglected.

In nature, the EI state is evidently rare. One obstacle
for creating an excitonic condensate is the far-off-equilibrium 
situation caused by optical excitation of excitons. But also in 
thermal equilibrium situations, EI states are expected 
to occur only under very particular circumstances, e.g., 
if conduction and valence bands are adequately nested~\cite{BBM02}.  
Actual materials experimentally studied from the viewpoint
of the EI are numbered. One recent example is quasi-one-dimensional
$\rm Ta_2Ni Se_5$ with highly polarizable Se, where 
angle-resolved photoemission spectra reveal an extreme valence-band 
top flattening indicating that the ground state might be viewed as 
an EI~\cite{WSTMANTKNT09}.
At present, the transition-metal dichalcogenide $1T$-$\rm TiSe_2$ 
seems to be the only candidate for a low-temperature
phase transition to the EI without the influence
of any external parameters other than the 
temperature.
Here the onset of an EI phase was invoked as driving force 
for the charge-density-wave transition~\cite{CMCBDGBA07,MCCBSDGBA09}. 
Semiconducting, pressure-sensitive mixed valence materials, 
such as ${\rm TmSe_{0.45}Te_{0.55}}$, are further candidates 
for exciton condensation. Fine tuning the excitonic level to 
the narrow 4$f$ valence band, a rather large number of
about $10^{20}-10^{21}$ $\rm cm^{-3}$  (small-to-intermediate sized) 
excitons can be created, which presumably condense
into an EI state, at temperatures below 20~K in the pressure 
range between 5 and 11 kbar~\cite{NW90,BSW91,WBM04}. 
Clearly, for these rather complex transition-metal/rare-earth compounds
with strong electronic correlations, 
simple effective-mass-model based theories will be too crude.  

The investigation of Falicov-Kimball-type models offers another
promising route towards the theoretical description of the EI 
scenario. In its original form, the Falicov-Kimball model~\cite{FK69,RFK70} 
contains two types of fermions: localized $f$ electrons and 
itinerant $c$ electrons with orbital energies $E_f$ and
$E_c$ , respectively. An on-site Coulomb interaction $U$ of both 
species determines the distribution of the electrons between 
the $f$ and $c$ sub-systems, and therefore may drive a valence transition, 
provided there is a way to establish $f$--$c$ coherence. 
At first glance this can only be achieved by including a hybridization 
of the $f$ and $c$ bands~\cite{KMM76,POS96a}. It has been shown, however, 
that a finite $f$ bandwidth, being certainly more realistic 
than entirely localized $f$ electrons, will also induce $f$--$c$ 
transitions~\cite{Ba02b,BGBL04}. The model with direct 
$c$--$c$ and $f$--$f$  hopping ($\propto t_{c/f}$),  
is sometimes called extended Falicov-Kimball model (EFKM), 
the Hamiltonian of which takes the form \begin{align}
H = &E_c \sum_i c_i^\dagger c_i^{} +  t_c \sum_{\langle i,j \rangle} 
c_i^\dagger c_j^{}  
+  E_f \sum_i f_i^\dagger f_i^{} 
+ t_f \sum_{\langle i,j \rangle} f_i^\dagger f_j^{} 
\nonumber \\
&+ U \sum_i  n_{ic}^{}n_{if}^{}\,.
\label{efkm}
\end{align}
Here $f_i^\dagger$ ($c_i^\dagger$) creates an $f$ ($c$) electron
at lattice site $i$, and $n_{if}=f_i^\dagger f_i^{}$ ($n_{ic}=c_i^\dagger 
c_i^{}$) are the corresponding number operators. 
Let us emphasize that the $f$ and $c$ bands 
involved have different parity~\cite{Ba02b}.
For $t_ft_c<0$ ($t_ft_c>0$), we may have a direct (indirect) band gap. 
For $t_f\equiv 0$ (dispersionless $f$ band), 
the local $f$ electron number is strictly 
conserved~\cite{SC08}.

In the past few years,  both the Falicov-Kimball model with 
hybridization~\cite{KMM76,POS96a,POS96b,Cz99,Fa99} and 
the EFKM~\cite{Ba02b,BGBL04,Fa08,SC08} have been studied in 
connection with the exciting idea of electronic ferroelectricity.  
The origin of electronic ferroelectricity is a spontaneously broken symmetry   
due to a non-vanishing $\langle c^\dagger f\rangle$ average,  
which causes finite  electrical 
polarizability without an external, interband-transition 
driving field. As $\langle c^\dagger f\rangle$ 
is basically an excitonic expectation value, 
indicating the pairing of $c$ electrons with $f$ holes, the
problem of electronic ferroelectricity is intimately 
connected with the appearance of an excitonic condensate.
Accordingly, the question whether the ground-state phase diagram of 
the EFKM exhibits an EI state has attracted
much attention.  By means of constrained path Monte Carlo (CPMC) 
techniques the $T=0$ phase diagram of the EFKM was determined 
in one and two dimensions in the strong and intermediate-coupling 
regimes~\cite{Ba02b,BGBL04}. In both cases a ferroelectric phase was detected.
A subsequent Hartree-Fock calculation shows that the mean-field 
phase diagram of the two-dimensional EFKM agrees even quantitatively with the 
CPMC data~\cite{Fa08}, supporting the applicability 
of Hartree-Fock and RPA schemes to three- and infinite-dimensional 
systems~\cite{Fa08,SC08,IPBBF08}.
Surprisingly, the more sophisticated scalar slave-boson theory failed 
to find the EI phase when the $f$ and $c$ orbitals are 
non-degenerate~\cite{Br08}. 

The continued controversy, regarding the existence of 
the EI phase in the EFKM, motivates us to re-examine 
the problem using an improved auxiliary boson approach
that ensures the rotational and gauge symmetries of the EFKM 
within a functional integral scheme. 
 
\section{Slave-Boson Theory}

\subsection{Slave-boson Hamiltonian}
\label{sec_sbho}
The extended Falicov-Kimball Hamiltonian~(\ref{efkm}) can be rewritten as an asymmetric Hubbard model
if  the orbital flavor $(f,c)$ is represented by  a pseudo-spin variable 
$(\sigma=\uparrow,\downarrow)$. Using the spinor representation  
\begin{equation}
\mathbf{a}_i= \left(\begin{matrix} a_{i\uparrow} \\ a_{i\downarrow} \end{matrix}\right)\,,\quad
\mathbf{a}_i^\dagger 	= (a_{i\uparrow}^\dagger, a_{i\downarrow}^\dagger)\,,\quad
\underline{t}=\left( \begin{matrix}	
\kappa & 0 \\ 0 & 1\end{matrix} \right)\,,
\label{spinor}
\end{equation}
where the vectors  $\mathbf{a}_i^{}$ ($\mathbf{a}_i^{\dagger}$) 
are built up by 
the fermion annihilation  (creation) operators  $a_{i\uparrow}^{(\dagger)}\equiv f_i^{(\dagger)}$ 
and $a_{i\downarrow}^{(\dagger)}\equiv c_i^{(\dagger)}$,
$H$ becomes 
\begin{equation}
H = \sum_{i\sigma} E_\sigma a_{i\sigma}^\dagger a_{i\sigma}^{} +\sum_{\langle i,j \rangle}\mathbf{a}_i^\dagger\; \underline{t}\;\mathbf{a}_j^{}
+ U \sum_i n_{i\downarrow} n_{i\uparrow}\,.	
\label{ashm}	
\end{equation}
In~Eq.~(\ref{spinor}), $\kappa=t_\uparrow / t_\downarrow$ gives the ratio of the $f$- and $c$-bandwidths
($t_\downarrow=1$ fixes the unit of energy). 
Obviously, the usual Hubbard model follows for
$E_\uparrow=E_\downarrow$ and $\kappa =1$. 
Without loss of generality we choose $E_\downarrow=0$ 
and $E_\uparrow\leq 0$ in what follows.  

Now the slave-boson representation of the EFKM  is constructed by 
replacing the fermionic Hilbert space  by an enlarged one of pseudo-fermionic and bosonic states.
The local states, representing the original physical states of~the EFKM~(\ref{efkm}) in the
enlarged Hilbert space in a one-to-one manner, can be created in the following way:  
\begin{align}
|0_i\rangle\;\;\rightarrow &\;\;e_i^\dagger |\mathrm{vac}\rangle\,,\\
|2_i\rangle\;\;\rightarrow &\;\;\tilde{a}_{i\uparrow}^\dagger 
\tilde{a}_{i\downarrow}^\dagger d_i^\dagger|\mathrm{vac}\rangle\,,\\
|\sigma_i \rangle\;\;\rightarrow &\;\; \sum_{\rho}
\tilde{a}_{i\rho}^\dagger p_{i\rho\sigma}^\dagger |\mathrm{vac}\rangle\,.
	\end{align}
The pseudo-fermions $\tilde{a}_{i\rho}$ satisfy anti-commutation rules
$\{\tilde{a}_{i\rho}^{},\tilde{a}^\dagger_{j\rho'}\}=\delta_{ij}\delta_{\rho\rho'}$,
while usual Bose commutation rules hold for the slave bosons: $[e_i^{},e_j^{\dagger}]= \delta_{ij}$, 
$[p_{i\rho_1\rho_2}^{},p^\dagger_{j\rho_3\rho_4}]=\frac{1}{2}\delta_{ij}\delta_{\rho_1\rho_4}\delta_{\rho_2\rho_3}$,
and $[d_i^{},d_j^{\dagger}]= \delta_{ij}$. The boson number operators,  
$e^\dagger_i e_i^{}$,  $2\, \mbox{Tr}\,{\underline{p}}^\dagger_i{\underline{p}}^{}_i $, 
and $d^\dagger_id_i^{}$ project on an empty, a singly occupied and a doubly occupied state, respectively.
By introducing a slave-boson matrix-operator for the case of single occupancy, $\underline{p}_i^{(\dagger)}$,
we adapt the spin-rotation-invariant slave-boson formulation of 
the Hubbard model~\cite{LWH89} (for a generalization to multi-orbital models see~\cite{LGKP07}), in order to avoid difficulties that may arise from the
scalar nature of the $p_\sigma$ bosons in approximative treatments~\cite{KR86,Br08}.  
The decomposition  
\begin{equation}
\underline{p}_i^{(\dagger)}=\frac{1}{2}\sum_{\mu=0}^3 
\underline{\tau}_\mu p_{i\mu}^{(\dagger)}
\end{equation}
into scalar (singlet) $p_{i0}^{(\dagger)}$  and vector (triplet) $\vec{p}_{i}^{\,(\dagger)}=(p_{ix}^{(\dagger)}, p_{iy}^{(\dagger)}, p_{iz}^{(\dagger)})$
components, where $\underline{\tau}_0$ is the unity 
matrix and $\underline{\vec{\tau}}$ the vector 
of Pauli spin matrices, is given as
\begin{align}
\underline{p}_i^{(\dagger)}=&\frac{1}{2}\left( \begin{matrix} 
p_{i0}^{(\dagger)} +p_{iz}^{(\dagger)} & p_{ix}^{(\dagger)} -ip_{iy}^{(\dagger)}
\\[0.1cm] p_{ix}^{(\dagger)} +ip_{iy}^{(\dagger)} & p_{i0}^{(\dagger)}-p_{iz}^{(\dagger)} 
\end{matrix}\right)\,.
	\end{align}

Of course, it is crucial to select out 
of the extended fermion-boson Fock space the physical 
states.  This can be achieved by imposing two sets of local constraints:
\begin{eqnarray}
C^{(1)}_i &=& e^\dagger_i e_i^{}\; 
	+ 2\, \mbox{Tr}\,
{\underline{p}}^\dagger_i{\underline{p}}^{}_i \,+ d^\dagger_id_i^{}\,- 1\,  =0\,,\label{c1}\\
\underline{C}^{(2)}_i &=& \tilde{\mathbf{a}}^{}_i \otimes 
  \tilde{\mathbf{a}}^\dagger_i +
	2 \, {\underline{p}}^\dagger_i{\underline{p}}^{}_i
        + d^\dagger_id_i^{}\,\underline{\tau}_0 -\underline{\tau}^{}_0 =0\,.
\label{c2}
\end{eqnarray}
$C^{(1)}_i$ expresses the completeness of the bosonic operators, i.e. each lattice site $i$ can 
only be occupied by exactly one boson. $\underline{C}^{(2)}_i$ relates the pseudo-fermion 
number to the number of $\underline{p}$ and $d$ bosons.

Correspondingly, the mapping of the physical electron operators into products 
of new pseudo-fermions and slave bosons in the  hopping term of $H$ is 
\begin{align}
\mathbf{a}_i \;\;\rightarrow &\;\; \underline{z}_i\, \mathbf{\tilde{a}}_i\,.
\label{a-mapping}
\end{align}
The choice of the hopping operators $\underline{z}_i$ is not unique. 
This arbitrariness can be
used, e.g., to reproduce, for the Hubbard model case, the correct free-fermion result 
at $U=0$  and the Gutzwiller result for any finite 
$U$ at the mean-field level, where 
the constraints~(\ref{c1}) and~(\ref{c2}) are fulfilled only on the average. 
This is guaranteed by choosing~\cite{LWH89,FW92b} 
\begin{equation}
{\underline{z}_i} = {\underline{L}_i}e^\dagger_i
        {{M}_i}{\underline{p}^{}_i} {\underline{N}_i}
        + {\underline{L}_i} {\tilde{\underline{p}}^{\dagger}_i}
        {{M}_i} d^{}_i {\underline{N}_i}\;
\label{E_zsri}
\end{equation}
\hspace*{1cm}with
\begin{eqnarray}
        {\underline{L}}^{}_i &=& [( 1 - d_i^\dagger d^{}_i )\underline{\tau}_0 - 
        2{\underline{p}}^\dagger_i{\underline{p}}^{}_i ]^{-1/2}\;,\\
        {\underline{N}_i} &=& [( 1 - e_i^\dagger e^{}_i )\underline{\tau}_0 - 
        2{\underline{\tilde{p}}}^\dagger_i{\underline{\tilde{p}}}^{}_i
        ]^{-1/2}\;,\\
        M^{}_i &=& [1 + e^\dagger_i e^{}_i + d^\dagger_i d^{}_i 
        + 2\, \mathrm{Tr}\, {\underline{p}}^\dagger_i{\underline{p}}^{}_i]^{1/2}\:,
\label{LMR_d}
\end{eqnarray}
and \mbox{$\tilde{p}^{\left(\dagger\right)}_{i\rho\rho '} =
\rho\rho ' p^{\left(\dagger\right)}_{i-\rho' -\rho}\;$}.
The Hubbard interaction term of $H$ can be bosonized via
\begin{align}
n_{i\uparrow} n_{i\downarrow} \;\;\rightarrow &\;\; d_i^\dagger d_i^{}\,.
\label{d_mapping}
	\end{align}
That is, the transformation to slave-boson fields results in a 
linearization of the interaction, 
and we end up with the EFKM Hamiltonian in the form 
\begin{equation}
H= \frac{E_\uparrow }{2}\sum_{i} \mathbf{\tilde{a}}_i^\dagger  
(\underline{\tau}_0+\underline{\tau}_z) 
\mathbf{\tilde{a}}_i
+ \sum_{\langle i,j\rangle} \mathbf{\tilde{a}}_i^\dagger 
\underline{z}_i^\dagger\, \underline{t}\, \underline{z}_j 
\mathbf{\tilde{a}}_j + U\sum_i d_i^\dagger d_i^{}\,.
\label{sbefkm}
\end{equation}
\subsection{Functional integral representation}
\label{sec_fire}
To proceed further, it is convenient to represent the grand 
canonical partition function of the constrained system~(\ref{sbefkm}),
$Z=\mathrm{Tr}\,\mathrm{e}^{-\beta(H-\mu N_e)}$, as an imaginary-time path integral over 
Grassmann fermionic and complex bosonic fields~\cite{NO88} 
\begin{eqnarray}
	Z &=&\int D [ \bar{\tilde{a}}_\rho, \tilde{a}^{}_\rho]\,
	D[e^\ast,e]\,D [p^\ast_\mu,p^{}_\mu]\,D[d^\ast,d]\,
	d[\lambda^{(1)}]\,d[\lambda^{(2)}_\mu]\nonumber\\&&\quad\quad\times\,
	\mathrm{e}^{-\int\limits_0^\beta\, d\tau L(\tau)}
\label{E_zusu1}
\end{eqnarray}
with the Lagrangian
\begin{eqnarray}
	L(\tau) &=& \sum_i\Big[- \lambda^{(1)}_i+
	e^\ast_i (\partial^{}_\tau+\lambda^{(1)}_i)e^{}_i\nonumber\\
	&&\quad\quad+2\,\mathrm{Tr}\,{\underline{p}^\ast_i}^T\big(
	 (\partial^{}_\tau + \lambda^{(1)}_i ) \underline{\tau}^{}_0
	- {\underline{\lambda}^{(2)}_i}^T
	\big){\underline{p}^{}_i}^T \nonumber\\[0.1cm]
        &&\quad\quad\quad\quad\quad+d^\ast_i(\partial^{}_\tau+\lambda^{(1)}_i+U- 
        \mathrm{Tr}\,{\underline{\lambda}^{(2)}_i})d^{}_i\Big]
\nonumber\\
&&+\sum_{i} \mathbf{\bar{\tilde{a}}}_i
 \big((\partial^{}_\tau -\mu) \underline{\tau}^{}_0 
+\underline{\lambda}_i^{(2)} +\tfrac{E_\uparrow}{2} (\underline{\tau}_0+\underline{\tau}_z)\big)\mathbf{\tilde{a}}_i \nonumber\\ && 
+\sum_{\langle i,j\rangle} \mathbf{\bar{\tilde{a}}}_i\,
\underline{z}_i^\ast \,\underline{t}\;\underline{z}_j\,\mathbf{\tilde{a}}_j\,.
\label{E_lagra}
\end{eqnarray}
Here $\beta=1/T$ is the inverse temperature and the time-independent Lagrange multipliers  
$\lambda_i^{(1)}$, 
$\underline{\lambda}_i^{(2)}=\lambda_{i0}^{(2)}\,\underline{\tau}_0^{}
+\vec{\lambda}_{i}^{(2)}\,\vec{\underline{\tau}}^{}$
are introduced to enforce the constraints via the integral 
representation of the $\delta$-function~\cite{HH87},
\begin{equation}
\delta \big[C^{(l)}\big] = \frac{\beta}{2\pi i}\int\limits_c^{c+2\pi i/\beta}\;
d\lambda^{(l)}\;e^{-\beta \lambda^{(l)} C^{(l)}} 
\label{E_df}
\end{equation}
(the path of the $\lambda$-integration is parallel to the imaginary axis and one finds $\lambda\in\mathbf{R}^+$
at the physical saddle point). 

Next, exploiting the gauge symmetry of the action, we perform 
local time-dependent phase transformations:
\begin{align}
e_i &\rightarrow e_i\, e^{-i\vartheta_i}\,, \\
d_i &\rightarrow d_i\, e^{-i\psi_i}\,,	\\
\underline{p}_i &\rightarrow \underline{p}_i\, e^{-i(\chi_{i0}\underline{\tau}_0-\chi_{iz}\underline{\tau}_z)} \,,\\
\mathbf{\tilde{a}}_i &\rightarrow \mathbf{\tilde{a}}_i \,e^{-i(\varphi_{i0}\underline{\tau}_0-\varphi_{iz}\underline{\tau}_z)}\,.
\label{etr}
\end{align}
Note that both the original as well as the transformed Bose fields are complex. By the transformation~(\ref{etr})
the kinetic contribution generates extra terms violating the 
$SO(2)\otimes U(1)^{\otimes 3}$ invariance of the model. Transforming
the Lagrange multipliers into real time-dependent  Bose fields,
\begin{align}
\lambda_i^{(1)} &\rightarrow \lambda_i^{(1)}+i\dot{\vartheta}_i\,, \\
\underline{\lambda}_i^{(2)} &\rightarrow e^{i(\chi_{i0}\underline{\tau}_0-\chi_{iz}\underline{\tau}_z)} \underline{\lambda}_i^{(2)}
e^{-i(\chi_{i0}\underline{\tau}_0-\chi_{iz}\underline{\tau}_z)} \nonumber \\
&\hspace{0.5cm}-i(\dot{\chi}_{i0}\underline{\tau}_0-\dot{\chi}_{iz}\underline{\tau}_z)
+i\dot{\vartheta}_i \underline{\tau}_0\,,
\end{align}
and, in addition, restricting the phase transformation to  $SO(2)\otimes U(1)^{\otimes 2}$ symmetry by 
\begin{align}
\psi_i &= 2\chi_{i0}-\vartheta_i\,, \\
\varphi_{i0} &= -\chi_{i0} +\vartheta_i\,,\\
\varphi_{iz} &= -\chi_{iz}\,,
\label{phaserestr}
\end{align}
the gauge invariance of the action is satisfied. We now make use of the gauge freedom to remove three 
phases of the Bose fields in radial gauge, where the fields are given as modulus times a phase factor:
\begin{align}
e_i &\rightarrow |e_i|\,e^{-i\tilde{\vartheta}_i}\,,\\
d_i &\rightarrow |d_i|\,e^{-i\tilde{\psi}_i}\,,\\
\underline{p}_i &\rightarrow \tfrac{1}{2} \sum_\mu |p_{i\mu}| \underline{\tau}_\mu e^{-i(\tilde{\chi}_{i0}\underline{\tau}_0
-\tilde{\chi}_{iz}\underline{\tau}_{z}})\,.
\end{align}
As a consequence, three bosons, e.g., $e_i(\tau)$, $p_{i0}(\tau)$, and $p_{iz}(\tau)$, can be taken as real-valued,
i.e., their kinetic terms, being proportional to the time derivates in 
Eq.~(\ref{E_lagra}), drop out due to the periodic boundary conditions imposed on Bose fields ($\phi_i(\beta)=\phi_i(0)$). However, the other three bosons $p_{ix}$, $p_{iy}$, and $d_i$  remain complex, 
\begin{align}
d_i &\rightarrow |d_i|\, e^{-i\psi_i}\;\; \mathrm{with}\; \psi_i=\tilde{\psi}_i-2\tilde{\chi}_{i0}+\tilde{\vartheta}_i \,,\\[0.1cm]
\underline{p}_i &\rightarrow \frac{1}{2}\left( \begin{matrix} p_{i0}+p_{iz} & (|p_{ix}|-i|p_{iy}|)e^{-2i\tilde{\chi}_{iz}} \\
(|p_{ix}|+i|p_{iy}|)e^{2i\tilde{\chi}_{iz}} & p_{i0}-p_{iz} \end{matrix} \right)\,.
\end{align} 
This has to be contrasted to the $SU(2)\otimes U(1)$ invariant Hubbard model ($t$-$J$ model),
where only one Bose field stays complex (all Bose fields become real)~\cite{DF94,DFB94,Fe96}. 

Using the familiar Grassman integration formula,
\begin{equation}
	\int D [ \bar{\tilde{a}}_\rho, \tilde{a}^{}_\rho]
	\mathrm{e}^{-\sum \bar{\tilde{a}}_\rho [-G^{-1}]_{\rho\rho'}
        \tilde{a}_\rho} =
	\mathrm{e}^{\;\mbox{\scriptsize Tr}\ln [-\underline{G}^{-1}]} \,,
\label{E_feint}
\end{equation}
the grand canonical partition function can be represented as a 
functional integral over Bose fields only, 
\begin{align}	
Z=\int &D[e] D [p^{}_0] D [p^{\ast}_x,p^{}_x] D [p^{\ast}_y,p^{}_y] D [p^{}_z] D[d^\ast,d]  \nonumber \\
&D[\lambda^{(1)}] D[\lambda^{(2)}_0] D[\vec{\lambda}^{(2)}] \,\mathrm{e}^{-S}
\label{E_zusta2}
\end{align}
with the effective bosonic action
\begin{eqnarray}
	S &=& \int\limits_0^\beta d\tau \bigg\{\sum_i
	\Big[- \lambda^{(1)}_i+
	\lambda^{(1)}_i e^2_i + \sum_\mu ( \lambda^{(1)}_i -
	\lambda^{(2)}_{i0} ) |p_{i\mu}|^2 \nonumber \\
	&&\qquad\quad\qquad- p^{ }_{i0}(\vec{p}^{\,\ast }_i+\vec{p}^{ }_i)
        \vec{\lambda}^{(2)}_{i} 
	-i\vec{\lambda}^{(2)}_i (\vec{p}^{\,\ast}_i\times \vec{p}^{ }_i)
          \nonumber \\[0.1cm] 
        &&\qquad\quad\qquad+(\lambda_i^{(1)} + U-2\lambda_{i0}^{(2)}) |d_{i}|^2 
         \nonumber \\
	&&\qquad\quad\qquad+ p^\ast_{ix}\partial_\tau p_{ix}+p^\ast_{iy}\partial_\tau p_{iy}+d_i^\ast \partial_\tau d_i^{}
	\Big]\bigg\}\nonumber \\ 
	       &&\;\;- \mbox{Tr}\, 
	\ln\Big\{-G^{-1}_{\langle ij\rangle,\rho \rho '}(\tau, \tau ')\Big\}\,, 
\label{E_seff}
\end{eqnarray}
where the inverse Green propagator is given by
\begin{align}
	G^{-1}_{\langle ij\rangle,\rho \rho '}(\tau, \tau ') =& 
	\Big[
	\big( -\partial^{ }_\tau + \mu - \lambda^{(2)}_{i0}\big)\delta^{ }_{\rho \rho '}
	\nonumber\\
        &- \tfrac{E_\uparrow}{2} (\underline{\tau}_0
        +\underline{\tau}_z)_{\rho \rho'}-\vec{\lambda}^{(2)}_{i} 
	\vec{\tau}_{\rho \rho '} 
	\Big]\delta^{ }_{i j}\,\delta (\tau - \tau ')\nonumber\\
	&-
	(\underline{z}^\ast_i\,\underline{t}\,\underline{z}^{}_j)^{}_{\rho \rho ', 
	\tau \tau '}(1-\delta^{ }_{i j})\,.
\label{E_Gij}
\end{align}
The trace in~Eq.~(\ref{E_seff}) extends over time, space, and spin variables.
 
The Hermitian $\underline{z}_i$ matrix can be brought into the form  
\begin{align}
\underline{z}_i	=
\left(\begin{matrix} |x_{i1}|^2 z_{i1}^{}+|x_{i2}|^2z_{i2}^{}&x_{i1}^{} y_{i1}^\ast z_{i1}^{} + x_{i2}^{} y_{i2}^\ast z_{i2}^{}\\
x_{i1}^\ast y_{i1}^{} z_{i1}^{}+ x_{i2}^\ast y_{i2}^{} z_{i2}^{}&|y_{i1}|^2 z_{i1}^{} + |y_{i2}|^2 z_{i2}^{} \end{matrix}\right) \, ,
\label{eqn:SUSB_zMatrix}
\end{align}
where
\begin{align}
\left(\begin{matrix} x_{i1} \\ y_{i1} \end{matrix} \right) =& \frac{1}{C_{i-}}\left(\begin{matrix} p_{ix}-ip_{iy}\\ p_i-p_{iz} \end{matrix}\right)\, ,\\
\left(\begin{matrix} x_{i2} \\ y_{i2} \end{matrix} \right) =& \frac{1}{C_{i+}}\left(\begin{matrix} p_{ix}-ip_{iy}\\-p_i-p_{iz} \end{matrix} \right)\, ,
\end{align}
are the eigenvectors of $\underline{p}_i$, $\underline{\tilde{p}}_i$ with
\begin{align}
p_i &= |\vec{p}_i\,|=\sqrt{|p_{ix}|^2 +|p_{iy}|^2 +p_{iz}^2 }\,,\\
C_{i\mp} &= [2 p_i (p_i\mp p_{iz})]^\frac{1}{2}\,,
\end{align}
and
\begin{eqnarray}
z_{i1}&=&\Big[(1-|d_i|^{2})-\tfrac{1}{2}\left(p_{i0}+ p_i \right)^2 \Big]^{-\frac{1}{2}}\nonumber \\
&&\times\tfrac{1}{\sqrt{2}}\Big[e_i \left( p_{i0}+ p_i \right)+
d_i \left( p_{i0}- p_i \right)\Big]\nonumber \\
&&\times\Big[ (1-e_i^2)-\tfrac{1}{2}\left(p_{i0}- p_i \right)^2 \Big]^{-\frac{1}{2}}\, , \label{eqn:SUSB-z1}\\
z_{i2}&=&\left[ (1- |d_i|^{2})-\tfrac{1}{2}\left(p_{i0}- p_i \right)^2 \right]^{-\frac{1}{2}}\nonumber \\
&  &\times\tfrac{1}{\sqrt{2}}\Big[e_i \left( p_{i0}- p_i \right)+
d_i \left( p_{i0}+ p_i \right)\Big]\nonumber \\
& &\times\Big[ (1-e_i^2)-\tfrac{1}{2}\left(p_{i0}+ p_i \right)^2 \Big]^{-\frac{1}{2}}\,.	\label{eqn:SUSB-z2}
\end{eqnarray}
Then we get 
\begin{eqnarray}
z_{i\uparrow\downarrow}^{}&=&x_{i1}^{} y_{i1}^\ast (z_{i1}^{}-z_{i2}^{})\,,	
\label{eqn:SUSB-nichtDiagZ1}\\
z_{i\downarrow\uparrow}^{}&=&x_{i1}^\ast y_{i1}^{} (z_{i1}^{}-z_{i2}^{})\,.	
\label{eqn:SUSB-nichtDiagZ2}	
\end{eqnarray}
We note that only for the half-filled band case 
($n_\uparrow+n_\downarrow=1$, $e_i=|d_i|$),  we find that 
$z_{i1}=z_{i2}=z_i$, i.e., 
\begin{equation}
\underline{z}_i	= z_i \underline{\tau}_0
\label{eqn:SUSB_zMatrix}
\end{equation}
becomes diagonal, and the matrix elements of the original Hamiltonian 
are reproduced by the slave-boson transformed model. That 
means,~Eq.~(\ref{E_zusta2}) with Eqs.~(\ref{E_seff}) to (\ref{eqn:SUSB_zMatrix}) 
provide an exact representation of the partition function for the EFKM at half-filling. 
By contrast, for the $SU(2)$-invariant Hubbard Hamiltonian with
$\kappa=1$ and $E_\uparrow=0$, the slave-boson 
representation of $Z$ holds exactly for all fillings.  

For the EFKM case with $t_\uparrow t_\downarrow <0$ 
(direct gap in the paraphase for large $|E_\uparrow |$), the EI order 
parameter $\Delta_{\perp}$ and the `Hartree shift' $\Delta_z$ are
respectively given as~\cite{SC08,Fa08,IPBBF08} 
\begin{align}\label{EIOP1}
\Delta_\perp &= \frac{U}{N}\sum_{i} \langle a_{i\downarrow}^\dagger a_{i\uparrow}^{} \rangle\,,\\
\Delta_z &= \frac{U}{N}\sum_{i\sigma} \sigma\, \langle a_{i\sigma}^\dagger 
a_{i\sigma}^{} \rangle\,.
\label{HS1}
\end{align} 
Using the constraints~(\ref{c2}) these relations can be expressed as 
functional averages:
\begin{align}\label{EIOP2}
\Delta_\perp &= \frac{U}{N} \sum_i \langle p_{i0}(p_{ix}-ip_{iy})\rangle\,,\\
\Delta_z &= 2\frac{U}{N}\sum_i \langle p_{i0} p_{iz}\rangle\,. 
\label{HS2}
\end{align}
\subsection{Saddle-point approximation}
\label{sec_spap}
The evaluation of Eq.~(\ref{E_zusta2}) is usually carried out by a loop expansion 
of the collective action~(\ref{E_seff}). At the first level of approximation, 
the bosonic fields are replaced by their time-averaged values, and
one looks for an extremum of the bosonized action with respect to the Bose and Lagrange multiplier fields 
$\phi_{i\alpha}=\big(e_i,p_{i0},\vec{p}_i,d_i,\lambda^{(1)}_i,
\lambda_{i0}^{(2)},\vec{\lambda}^{(2)}_i\big)$:
\begin{eqnarray}
        \frac{\partial S}{\partial\phi_{i\alpha}}\stackrel{!}{=}0
        \;\;\;\leadsto\;\;\;
        \bar{S} = {S}
        \Big|_{{\phi}_{i\alpha}=\bar{\phi}_{i\alpha}}\,. 
\label{E_AblSeff}
\end{eqnarray}

The physically relevant saddle point $\{\bar{\phi}_{i\alpha}\}$ is
determined to give the lowest free energy (per site), 
\begin{equation}
        \bar{f}= \bar{\Omega}/N + \mu n\,,
\label{barf}
\end{equation}
where, at given mean electron density $n=n_\uparrow +n_\downarrow$, 
the chemical potential $\mu$
is fixed by the requirement 
\begin{equation}
n = - \frac{1}{N}\frac{\partial \bar{\Omega}}{\partial \mu}\,.
\end{equation}
$\bar{\Omega} = \bar{S} / \beta $ 
denotes the grand canonical potential.
Clearly, an unrestricted minimization of the free energy is impossible for an infinite system,
even within the static approximation. Focusing on the possible existence of the EI phase, we consider only 
uniform solutions hereafter: $\{\bar{\phi}_{i\alpha}\}=\{\bar{\phi}_{\alpha}\}$.
Note that the inclusion of a charge-density-wave 
phase is straightforward, 
e.g., by adapting the two-sublattice slave-boson 
treatment worked out for  the Peierls-Hubbard model~\cite{FDB92,TFDB94,Fe96}.

Examining a tight-binding direct-band-gap situation in three dimensions,
we have 
\begin{equation}
\underline{z}^\ast\;\underline{\varepsilon}_{\vec{k}}\;\underline{z}=
z^2 \gamma_{\vec{k}} \,\underline{t}
\label{z2u}
\end{equation}
with
\begin{equation}
z^2= \frac{2p_0^2d^2}{[1-d^2-\frac{1}{2}(p_0+p)^2] 
[1-d^2-\frac{1}{2}(p_0-p)^2]} 
\end{equation}
and 
\begin{equation}
\gamma_{\vec{k}} =-2  [\cos k_x +\cos k_y +\cos k_z ]\,.
\label{gammak}
\end{equation}
The trace in~Eq.~(\ref{E_seff}) can be easily performed in the momentum-frequency-domain after diagonalizing 
the propagator in pseudo-spin space. 
Then the free energy functional takes the form
\begin{eqnarray}
f[\phi_\alpha]&=&\lambda^{(1)} (e^2+p_0^2+p^2+d^2-1) 
\nonumber\\
&&-2\lambda_\perp^{(2)} p_0 p_\perp -2\lambda_z^{(2)} p_0 p_z +Ud^2
\nonumber\\
&&+\frac{1}{\beta N}\sum_{\vec{k}\nu} \ln \left[ 1-n_{\vec{k}\nu}\right] +\tilde{\mu} n\,,
\end{eqnarray}
where
\begin{equation}
n_{\vec{k}\nu}= [\exp \{ \beta(E_{\vec{k}\nu}-\tilde{\mu}) \}+1]^{-1}
\label{nknu}
\end{equation}									
holds with the quasiparticle energies $(\nu=\pm)$
\begin{eqnarray}\label{Eknu}
E_{\vec{k}\nu}&=&\tfrac{1}{2}[E_\uparrow+(\kappa+1)z^2\gamma_{\vec{k}}]\\&& +\nu 
\sqrt{\tfrac{1}{4}[E_\uparrow+2\lambda^{(2)}_z+(\kappa-1)z^2\gamma_{\vec{k}}]^2+(\lambda^{(2)}_\perp)^2}\,.\nonumber	
\end{eqnarray}
Here we have introduced $\lambda_\perp^{(2)}=\pm\sqrt{(\lambda^{(2)}_x)^2
+(\lambda^{(2)}_y)^2}$, $p_\perp=\mp\sqrt{p_x^2+p_y^2}$,  and $\tilde{\mu}=\mu-\lambda_0^{(2)}$.

Requiring that $f$ becomes stationary with respect to the variation of the $\phi_\alpha$
we obtain the following set of saddle-point equations: 
\begin{align}
	\lambda_z^{(2)}=&\frac{1}{2} \frac{p_z}{p_0}  
\left( \frac{z^2}{2d^2}-\frac{1}{p_0^2-p^2} \right)\,z^2 I\,,\label{lambda2z}\\	
\lambda_\perp^{(2)}=&\frac{1}{2} \frac{p_\perp}{p_0} 
\left( \frac{z^2}{2d^2}-\frac{1}{p_0^2-p^2} \right)\,z^2 I\,,\label{lambda2perp}\\
p_0p_z =&	\frac{1}{2}\frac{1}{N} \sum_{\vec{k}\nu} 
\nu m_{\vec{k}} n_{\vec{k}\nu}\,,\label{p0pz}\\
p_0p_\perp =&\frac{1}{2}\frac{1}{N} \sum_{\vec{k}\nu} \nu M_{\vec{k}} n_{\vec{k}\nu}\,,\label{p0pperp}\\	
p_0^2=&\frac{1}{2}+\frac{1}{2} \sqrt{(1-z^2)(1-4p_0^2p^2)} \,,\\
d^2 =& \frac{1}{2z^2}\biggl(z^2 (2-p_0^2-p^2)+2p_0^2 \nonumber \\
&-2p_0 \sqrt{z^2 (2-p_0^2-p^2)+z^4 p^2+p_0^2} \biggr)\,,\\
U+2&\lambda_\perp^{(2)}\frac{p_\perp}{p_0}+2\lambda_z^{(2)}\frac{p_z}{p_0}=
\frac{2 d^2-p_0^2+z^2p^2}{2p_0^2d^2}\,z^2 I
\label{corrvectorsb}
\end{align}
with
\begin{align}\label{I}
I= &(\kappa+1)\frac{1}{N} \sum_{\vec{k}\nu} \gamma_{\vec{k}}
n_{\vec{k}\nu}+ (\kappa -1) \frac{1}{N} \sum_{\vec{k}\nu} \nu m_{\vec{k}} \gamma_{\vec{k}}n_{\vec{k}\nu}\,,\\
m_{\vec{k}}\!=&\frac{ E_\uparrow+2\lambda_z^{(2)} + (\kappa-1)z^2\gamma_{\vec{k}}}{\sqrt{(E_\uparrow+2\lambda_z^{(2)} +(\kappa-1)z^2\gamma_{\vec{k}})^2
+(2\lambda_\perp^{(2)})^2}} \,,\label{m_k}\\
M_{\vec{k}}=&\frac{2\lambda_\perp^{(2)}}{\sqrt{(E_\uparrow+2\lambda_z^{(2)} +(\kappa-1)z^2\gamma_{\vec{k}})^2+(2\lambda_\perp^{(2)})^2}}\,.\label{M_k}
\end{align}
The EI order parameter~(\ref{EIOP1}) and Hartree shift~(\ref{HS1}) become 
\begin{align}
\label{EIOP3}
\Delta_\perp &= Up_0 p_\perp\,,\\
\Delta_z &= 2 Up_0 p_z\,.
\label{HS3}
\end{align}
\subsection{Zero temperature: BI to EI transition}
\label{sec_bieitr}
At zero temperature, the EFKM exhibits a trivial band insulator (BI) phase  
of a completely filled $f$ (empty $c$) band ($n=1$, $E_{\uparrow}\leq 0$), 
provided the Hartree gap is finite: 
\begin{align}
\Delta_{H}(T=0)=|E_\uparrow+2\lambda^{(2)}_z|-6(|\kappa| +1) >0\,.
\label{deltahartree}
\end{align}
That is to say, in the BI phase we have $d^2=0$, $n_\uparrow=1$, $n_\downarrow=0$,
and in no way $f$-$c$ coherence can develop: $p_\perp=\lambda_\perp^{(2)}=0$.
Then $m_{\vec{k}}=-1$ for all $\vec{k}$, and $n_{\vec{k}-}=1$, $n_{\vec{k}+}=0$
result from~Eq.~(\ref{nknu}) with~Eq.~(\ref{Eknu}). The constraint~(\ref{c2}) gives, together with~Eq.~(\ref{p0pz}),  $p_0=p_z=1/\sqrt{2}$, leading to $z^2=1$, as for a non-interacting system. At the same time, $I=0$ according to
Eq.~(\ref{I}), and the correlation equation (\ref{corrvectorsb}) reduces
to $U+2\lambda^{(2)}_z=0$, which gives $\Delta_z=-2\lambda^{(2)}_z$ for the BI.

Looking for an instability of the BI towards an EI state, we find 
from~Eq.~(\ref{lambda2perp}) that $\lambda_\perp^{(2)}=-\Delta_\perp$ near 
the critical Coulomb interaction $U_{c2}$. Multiplying Eq.~(\ref{p0pperp}) by $U$,
for $\Delta_\perp\neq 0$, we get the $T=0$ gap equation  
\begin{align}
1 =&& \frac{1}{2}\frac{U}{N} \sum_k \frac{n_{\vec{k} -} - n_{\vec{k} +}}
{\sqrt{\tfrac{1}{4}[E_\uparrow - \Delta_z +(\kappa-1) \gamma_{\vec{k}}]^2 
+\Delta_\perp^2}} \,,
\end{align}
which agrees with the Hartree-Fock result~\cite{IPBBF08}.
As a consequence, our $SO(2)$-invariant slave-boson approach 
reproduces the BI-EI phase boundary of the 
EFKM Hartree-Fock ground-state phase diagram~\cite{SC08,Fa08}.  
At least for the 2D case it has been demonstrated that this
phase boundary agrees almost perfectly with that obtained
by the CPMC method~\cite{Fa08,BGBL04}. This also applies
to our $SO(2)$-invariant slave-boson approach, e.g.,
for the 2D EFKM with $\kappa=-0.3$ and $U=2$, we obtain the 
critical value  $|E_{\uparrow,c}|=3.23 (3.26)$ for the EI--BI transition, 
using the 2D tight-binding (square) density of states, in comparison with 
$|E_{\uparrow,c}^{\text{CPMC}}|=3.29$ (cf. Fig.~3 in Ref.~\onlinecite{BGBL04}). 

\subsection{Scalar slave-boson approach}
\label{sec_ssba}
If one contrariwise adopts the scalar slave-boson theory~\cite{KR86} 
by introducing only four auxiliary bosonic fields per 
site, $e_i$, $p_{i\sigma}$, and $d_i$,
where $p^\dagger_{i\uparrow}p^{}_{i\uparrow}$ 
($p^\dagger_{i\downarrow}p^{}_{i\downarrow}$)  projects on a singly occupied
$f$- ($c$-) electron site $i$, the $\underline{p}_i$-matrix becomes 
diagonal:
\begin{align}
\underline{p}_i=&\frac{1}{2}\left( \begin{matrix} p_{i0} +p_{iz} & 0 \\ 0 & 
p_{i0}-p_{iz} \end{matrix}\right)\,. 
\end{align}
That means, the `spin-flip' terms in Eq.~(\ref{c2}), 
\begin{align}
\tilde{a}_{i\uparrow}^\dagger \tilde{a}_{i\downarrow}=&2p_{i\uparrow\downarrow}^\dagger p_{i\uparrow\uparrow}+2p_{i\downarrow\downarrow}^\dagger p_{i\uparrow\downarrow}			\label{eqn_SUSB_NB3} \\
\tilde{a}_{i\downarrow}^\dagger \tilde{a}_{i\uparrow}=&2p_{i\uparrow\uparrow}^\dagger p_{i\downarrow\uparrow} + 2p_{i\downarrow\uparrow}^\dagger
p_{i\downarrow\downarrow}		\label{eqn_SUSB_NB4} \,,
\end{align}
do not occur. As these terms, in view of Eqs.~(\ref{EIOP1}) 
and~(\ref{EIOP2}),  are essential 
for the formation of an excitonic insulator, the scalar 
slave-boson approach fails to describe the EI phase, at least for 
finite orbital-energy difference $E_\uparrow\neq 0$.    
At the (uniform) saddle-point level of approaximation, 
within scalar slave-boson theory, we find
the band-renormalization factor
\begin{align} 
z^2 &= \frac{d^2 (p_\uparrow + p_\downarrow)^2 }{n_\uparrow n_\downarrow}
\label{zscsb}
\end{align}
with $p_\uparrow=\sqrt{2}p_{\uparrow\uparrow}$ 
($p_\downarrow=\sqrt{2}p_{\downarrow\downarrow}$), and 
the correlation equation~(\ref{corrvectorsb}) simplifies to
\begin{align}
U &= z^2 \left( \frac{1}{p_\uparrow p_\downarrow} -\frac{1}{d^2} \right) \frac{1}{N} \sum_{\vec{k}} \bigl( \kappa \gamma_{\vec{k}} n_{\vec{k} \uparrow} 
+\gamma_{\vec{k}} n_{\vec{k} \downarrow} \bigr)\,.
\label{corrssb}
\end{align}
\section{Numerical Results}
In the numerical evaluation of the self-consistency loop~(\ref{barf})--(\ref{M_k})
we proceed as follows: at given model parameters $E_f$, $\kappa$,
$U$, and fixed total particle density $n=n_f+n_c=1$, we solve the 
finite-temperature saddle-point equations for the 
slave-boson and Lagrange parameter fields $\{\bar{\phi}_\alpha\}$ together with 
the equation for the renormalized chemical potential $\tilde{\mu}$ 
using an iteration technique. Thereby $\vec{k}$-summations were transformed 
into energy integrals, introducing the (tight-binding) density of 
states for the simple cubic lattice. Convergence is assumed to be achieved
if all quantities are determined with a relative error less than $10^{-6}$.
Our numerical scheme allows for the investigation of different metastable
states corresponding to local minima of the variational  
free-energy functional. Of course, we will always obtain a homogeneous,
translational invariant solution without spontaneous (polarization) 
exciton formation. In this case the $f$ and $c$ bands are simply shifted 
by $2\lambda_z^{(2)}$, leading to a gapped band structure at large enough $U$.  
Besides these simple (semi-) metallic and BI phases, the $T=0$ 
Hartree-Fock ground-state phase diagram of the half-filled EFKM exhibits 
two symmetry-broken states~\cite{Fa08,SC08}: the anticipated EI and  
a charge-density-wave phase.  At $E_f=E_c$ (degenerate orbitals), the 
charge-density-wave ground state is stable for all values of $\kappa$. 
It becomes rapidly suppressed, however, for $E_f\neq E_c$ (non-degenerate orbitals), 
in particular, if the $c$ and $f$ bandwidths are comparable~\cite{Fa08}.  
As we are interested in the (uniform) EI phase only, we have 
confined our slave-boson approach to spatially uniform saddle points. 
With respect to charge-density-wave formation this will be uncritical 
for the parameter values studied in the following.  

\begin{figure}[t]
  \centering 
  \includegraphics[width=\linewidth,clip]{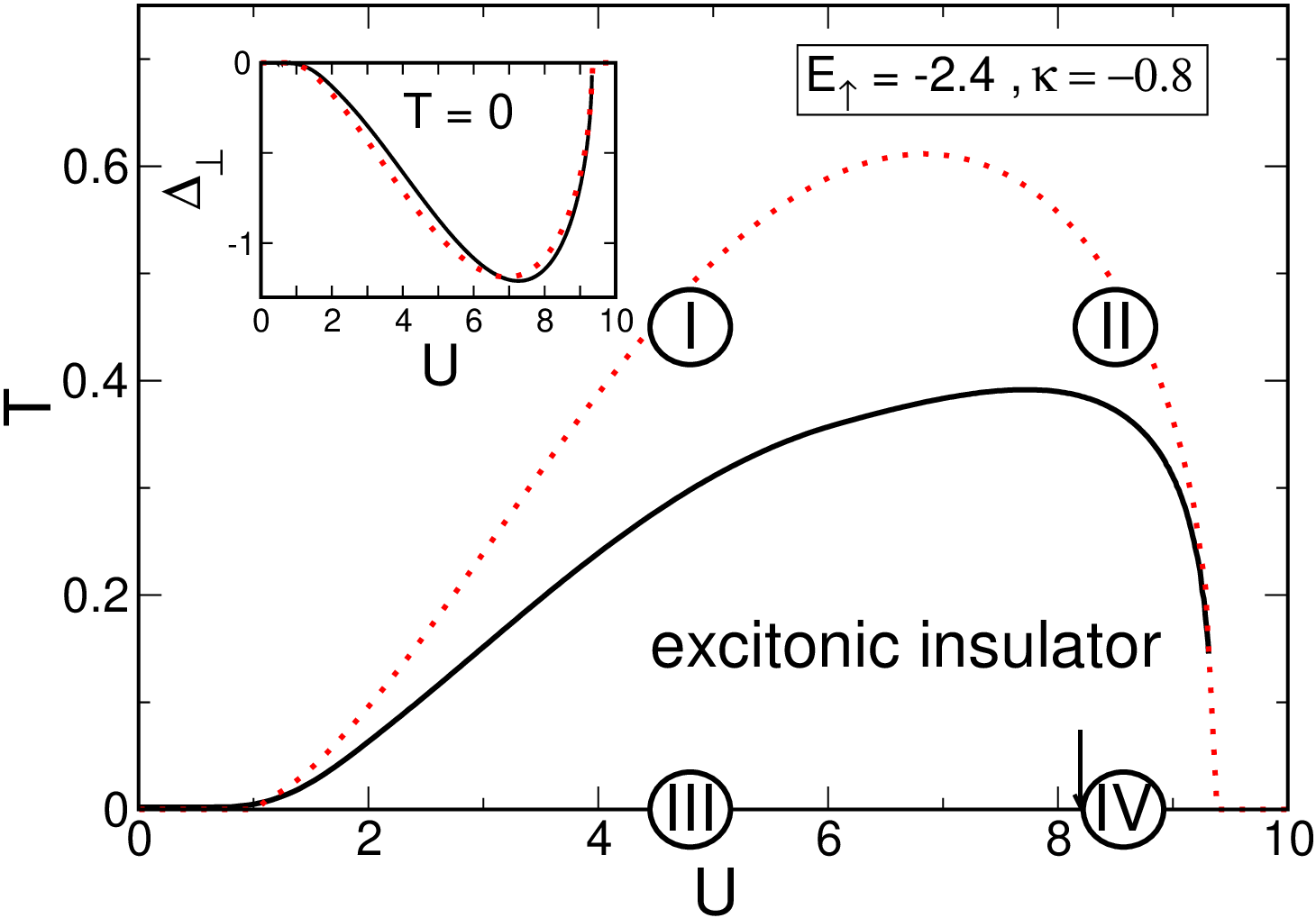}
  \caption{(Color online) Stability region of the EI phase in the 
  3D half-filled EFKM  (the arrow marks the critical 
  coupling where the Hartree gap~(\ref{deltahartree}) opens). 
  The inset shows the order parameter at zero temperature. 
  Red dotted curves give the Hartree-Fock results for comparison.
\label{fig_pd}}
\end{figure}

Figure~\ref{fig_pd} gives the slave-boson phase boundary of the EI 
in the $U$-$T$ plane, calculated for $E_\uparrow=-2.4$ and $\kappa=-0.8$.
Most notably, we obtain a stable EI solution for the non-degenerate
band case, which has to be contrasted with the result of the scalar slave-boson
approach~\cite{Br08}. 

Let us first discuss the $T=0$ data.   
Here the  numerical semimetal-EI and EI-BI transition points at small and 
large Coulomb interaction, $U_{c1}\simeq 0.74$ and $U_{c2}\simeq 9.3$, 
respectively, agree with the Hartree-Fock results (see the dotted curve). 
The latter was proved analytically in Sec.~\ref{sec_bieitr}. 
The inset gives the $U$-dependence of
the EI order parameter at $T=0$. 
For $U_{c1}\leq U\leq U_{c2}$, $\Delta_\perp$ 
only slightly deviates from the corresponding Hartree-Fock curve. 

\begin{figure}[t]
  \centering 
\includegraphics[width=0.48\linewidth,clip]{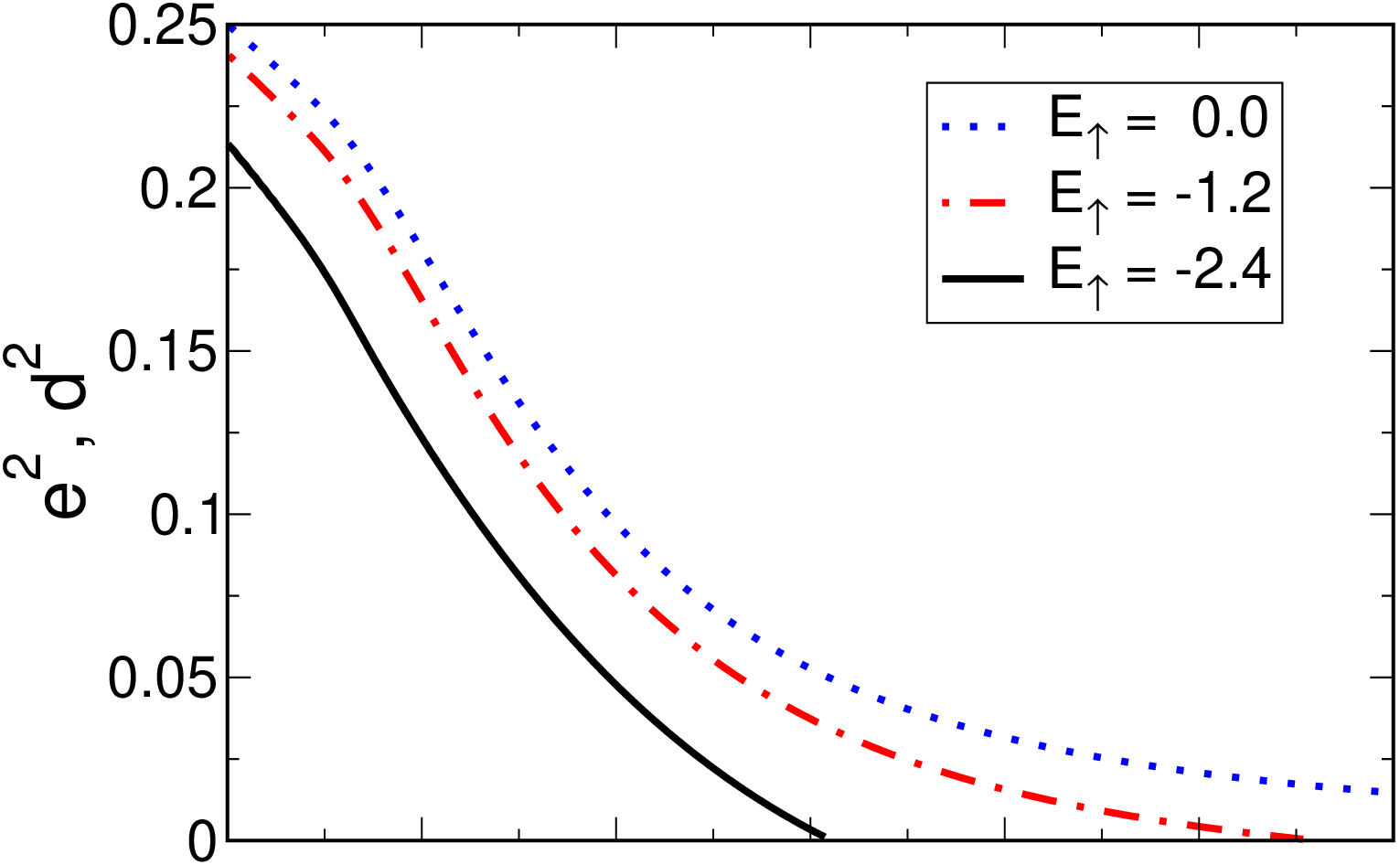}\hspace*{0.2cm}
\includegraphics[width=0.48\linewidth,clip]{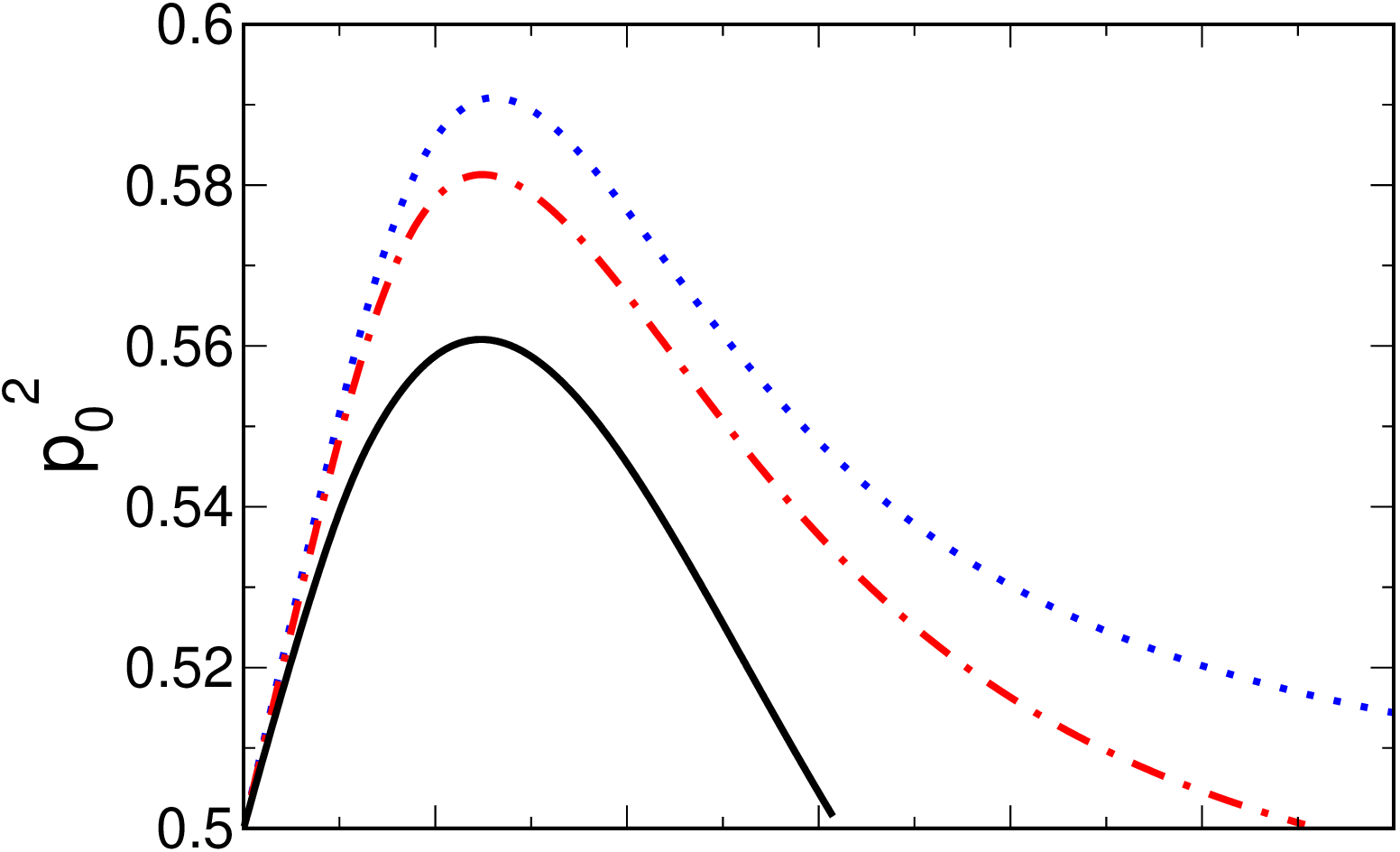}\\[0.1cm]
\includegraphics[width=0.48\linewidth,clip]{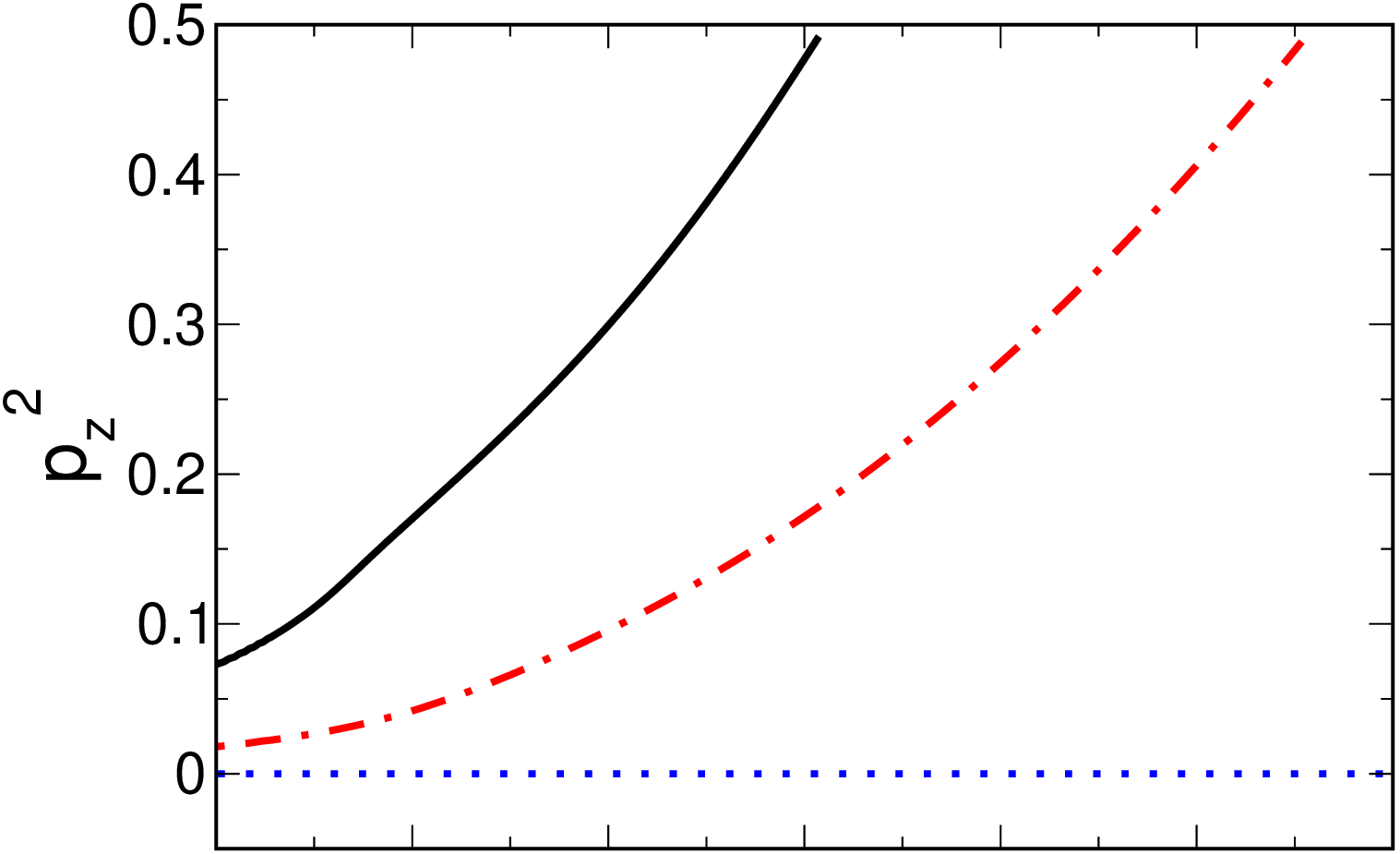}\hspace*{0.2cm}
\includegraphics[width=0.48\linewidth,clip]{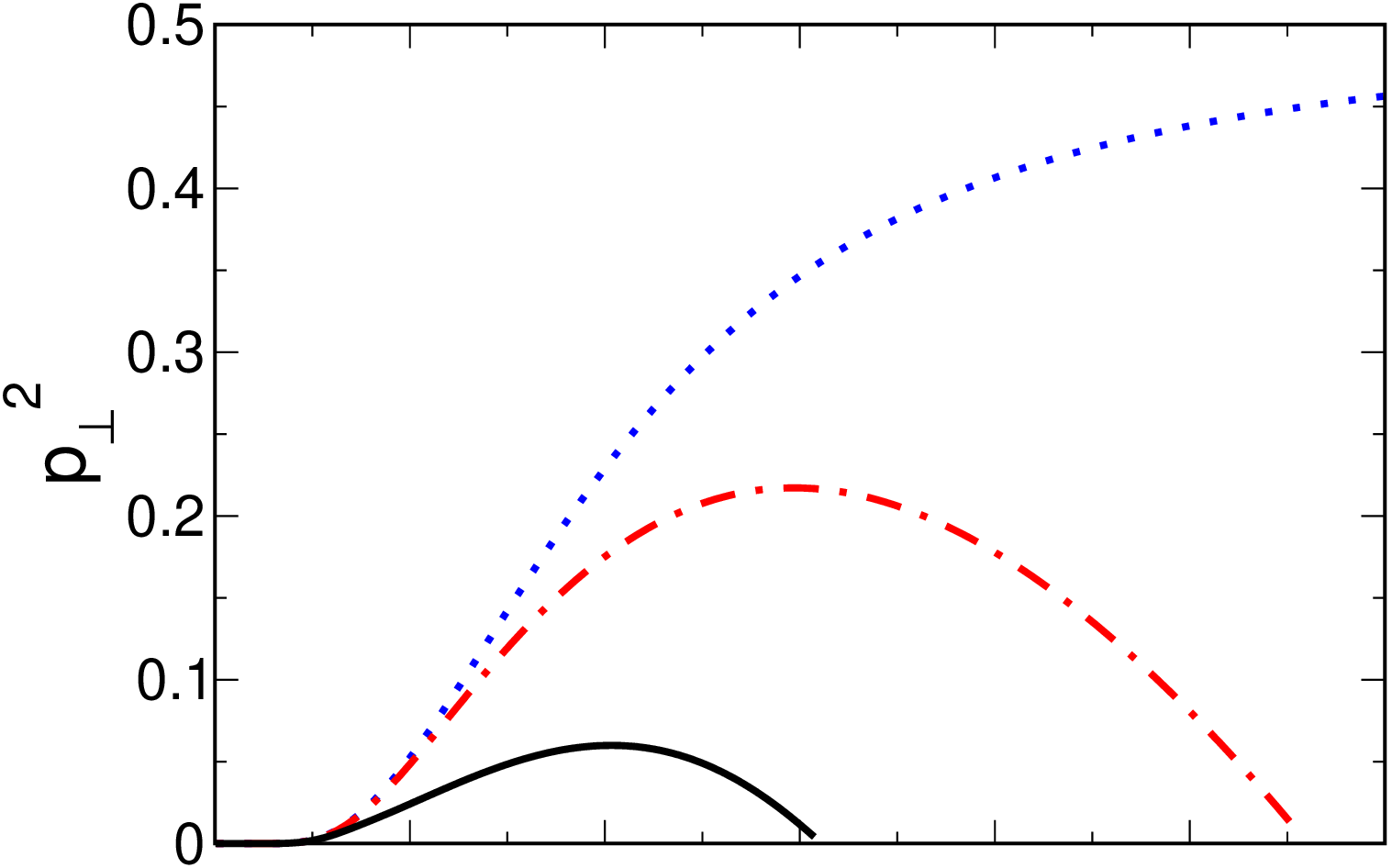}\\[0.1cm]
\hspace*{0.1cm}\includegraphics[width=0.48\linewidth,clip]{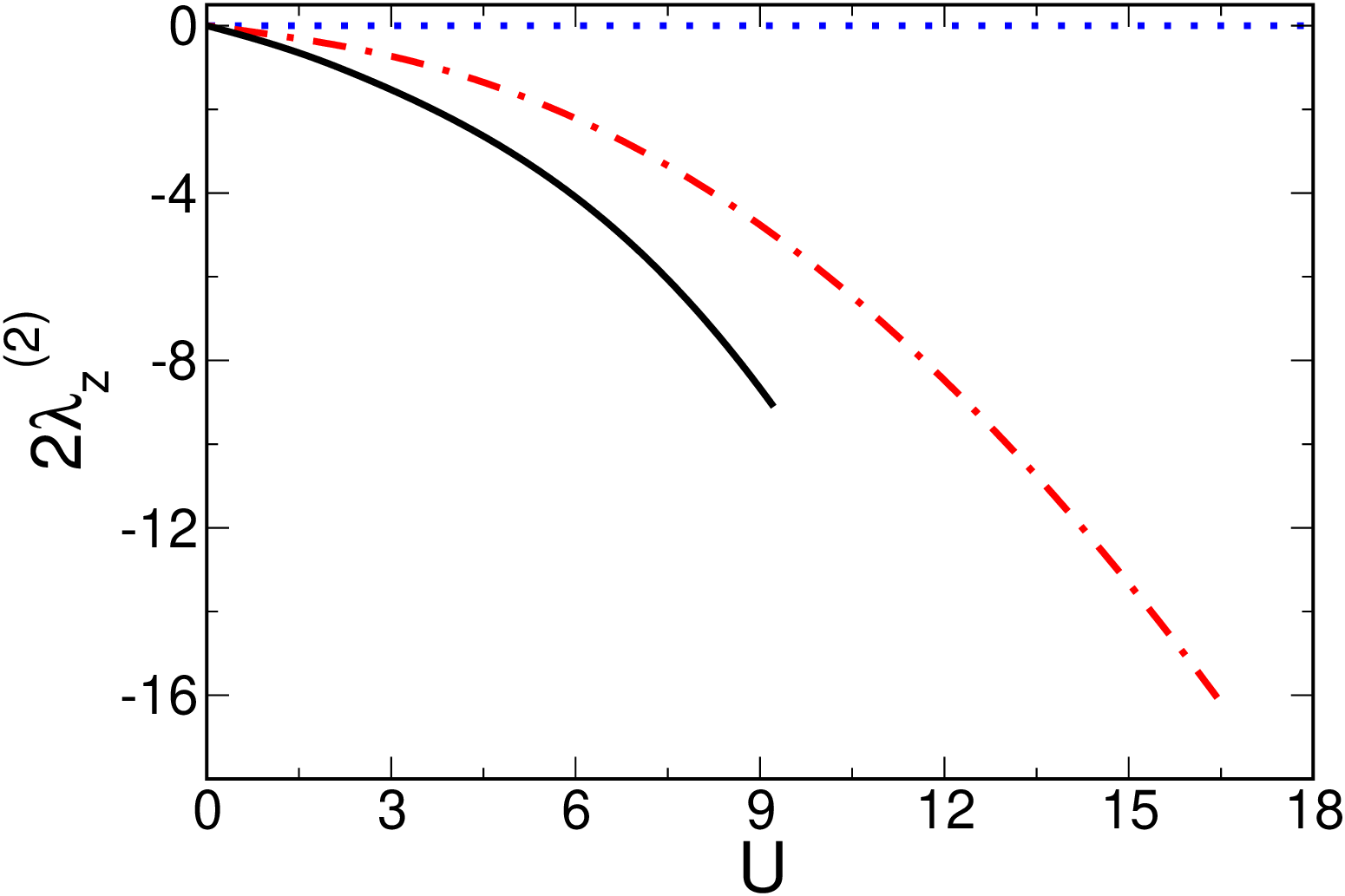}\hspace*{0.3cm}
\includegraphics[width=0.46\linewidth,clip]{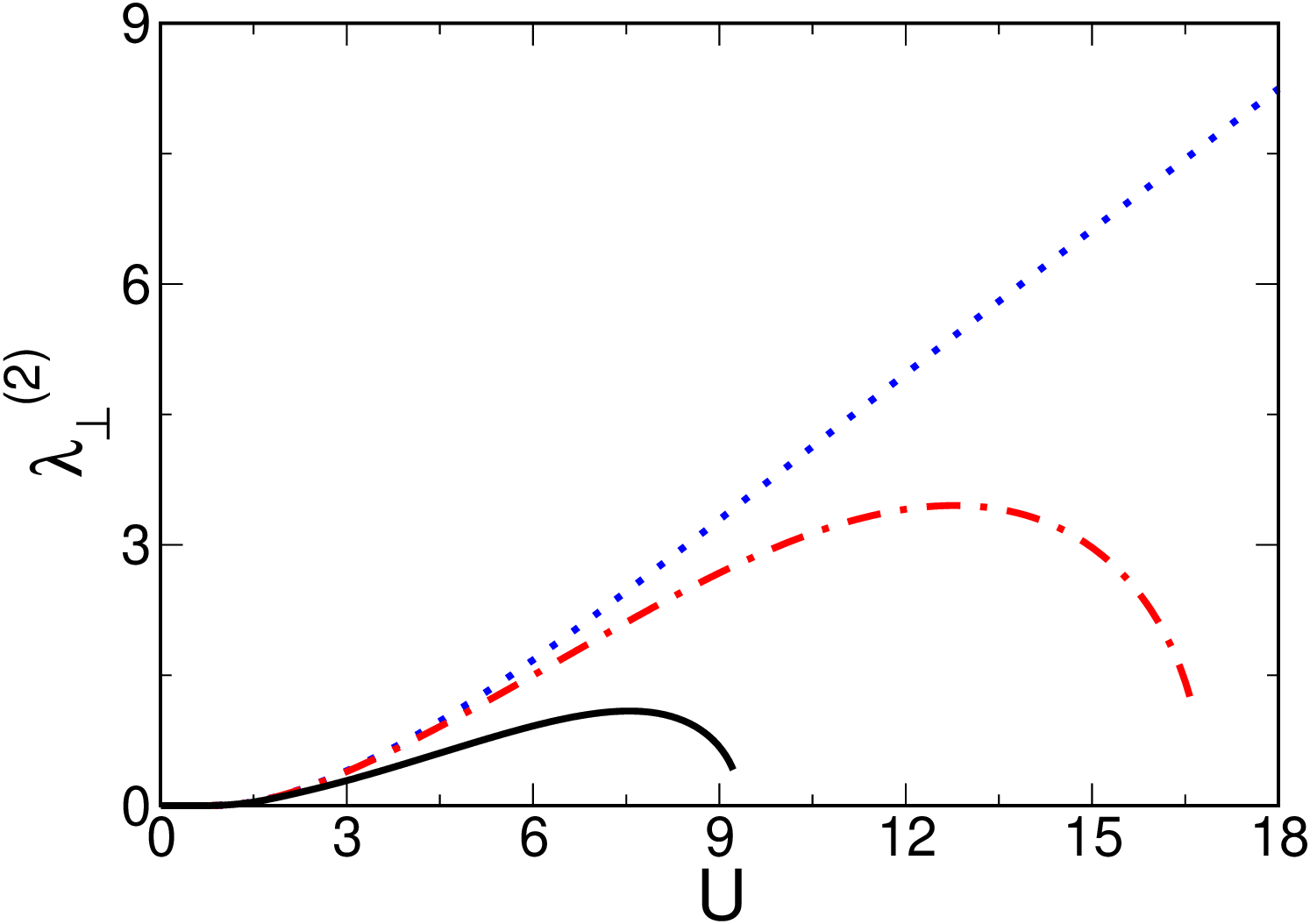}
  \caption{(Color online) $U$-dependence of slave-boson fields and 
 Langrange parameters for different $E_\uparrow$ at $T=0$ ($\kappa=-0.8$).}
  \label{fig_sbt0}
\end{figure}
The variation of the other bosonic fields is depicted in    
Fig.~\ref{fig_sbt0}, where the solid curves belong 
to the parameter values used in Fig.~\ref{fig_pd}. We see 
that the number of empty and double-occupied sites, $e^2$ and $d^2$,
is equal and goes to zero at the EI-BI transition, where 
we have $p_0^2=p_z^2=1/2$ at singly occupied sites. 
Non-vanishing values of $p_\perp^2$ and $\lambda_\perp^{(2)}$
indicate an EI state, which demonstrates the importance
of the (transverse) `spin-flip' processes for the formation and
maintenance of $f$-$c$ coherence. The slave-boson band shift 
$|2\lambda_z^{(2)}|$ in Eq.~(\ref{Eknu}) increases with increasing $U$ (just as the 
Hartree shift). Obviously, the area of the EI phase 
is enlarged if one reduces the splitting of the $f$ and $c$ band 
centers (cf. the red dot-dashed lines). We include the data for the 
metastable EI solution at $E_\uparrow =0$ (as discussed above, in this 
case the charge-density-wave state will win), in order to show that $d^2$
and $e^2$ stay finite for all $U$. That means, for the orbital-degenerate
EFKM ($\lambda_z^{(2)}=\tilde{\mu}=0$), our $SO(2)$-invariant slave-boson 
scheme will not give the (artificial) transition into an insulating 
Brinkmann-Rice-like correlated-insulator state~\cite{BR70}. This 
transition is a well-known shortcoming
of the scalar slave-boson approach to the Hubbard model~\cite{KR86} and
has been also observed applying the scalar slave-boson theory to 
the EFKM~\cite{Br08}. The effect becomes even more apparent by comparing the
variation of the slave-boson band-renormalization factors $z^2$.

\begin{figure}[t]
  \centering 
\includegraphics[width=0.49\linewidth,clip]{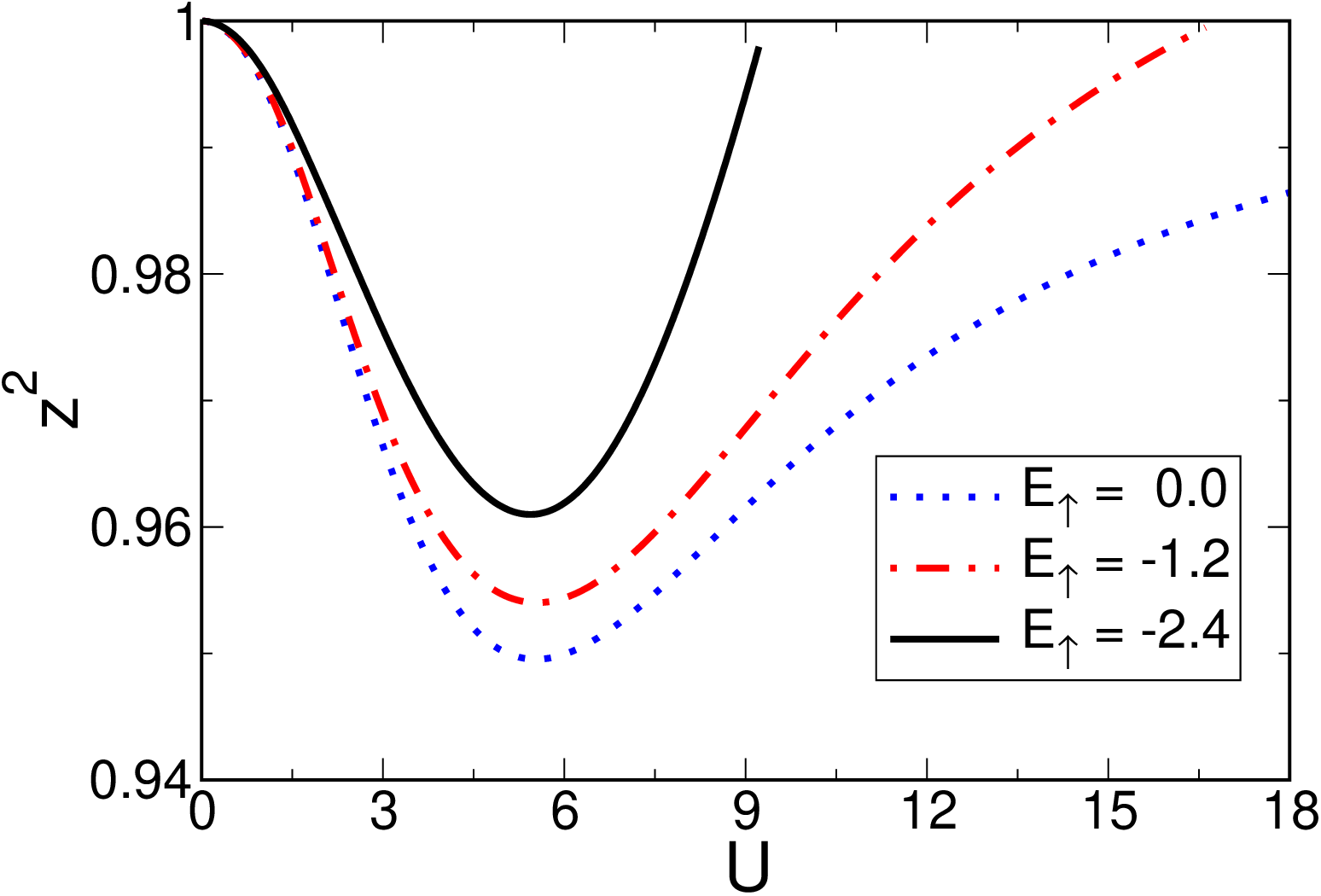}\hspace*{0.2cm}
\includegraphics[width=0.49\linewidth,clip]{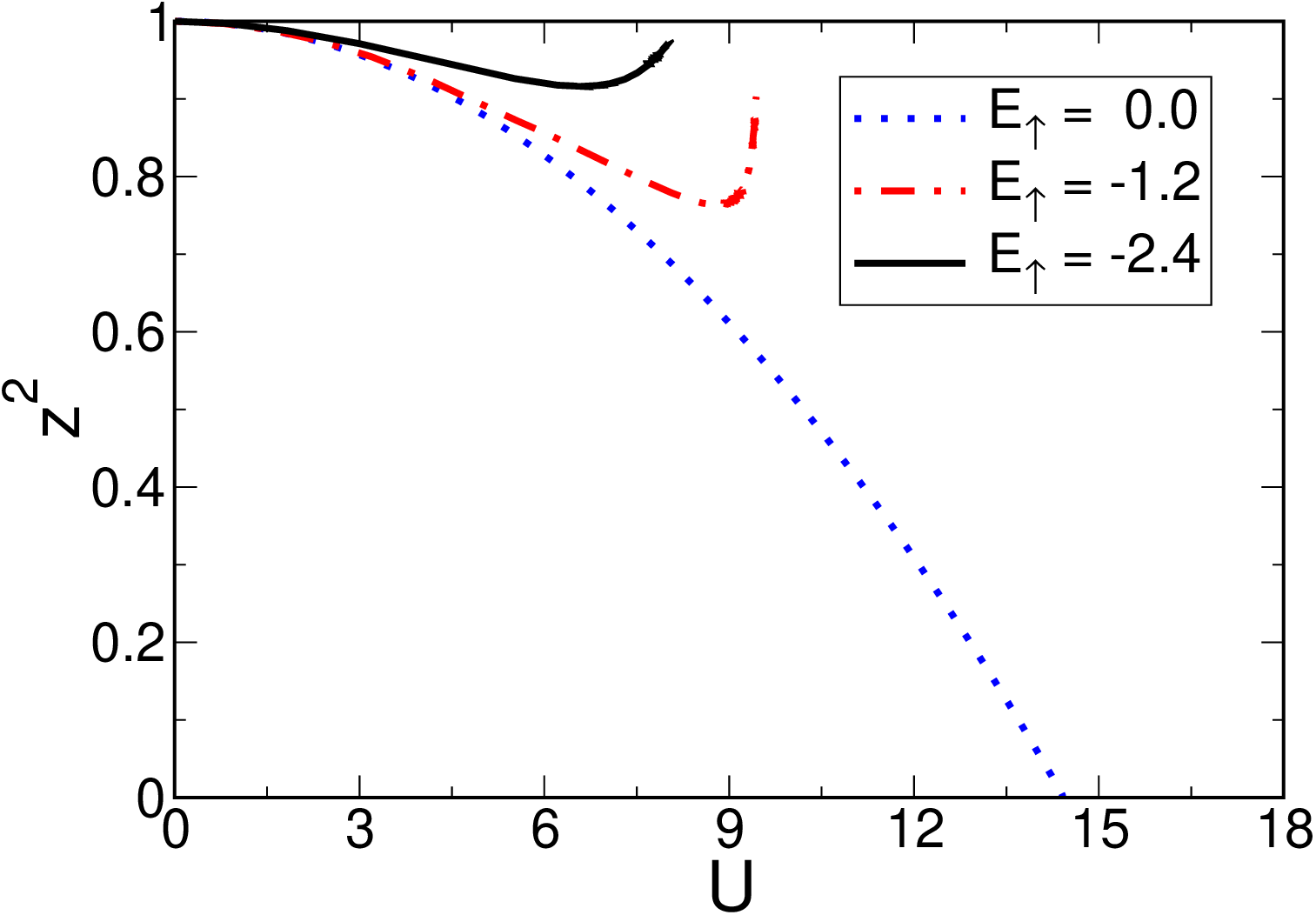}
  \caption{(Color online) Band-renormalization factors at $T=0$ within 
$SO(2)$-invariant (left-hand panel) and scalar
  (right-hand panel) slave-boson theory. Again, $\kappa=-0.8$.}
  \label{fig_zfactor}
\end{figure}

Figure~\ref{fig_zfactor} shows that $z^2$ vanishes within the scalar 
slave-boson theory (right-hand panel) for $E_\uparrow=0$ at a critical 
interaction strength ($U_{\rm BR}\simeq 14.5$), indicating the localization of 
charge carriers, whereas in our theory the bandwidth will be only slightly renormalized 
at this point (see left-hand panel). Interestingly, the band 
renormalization is rather small in the EI phase as well (cf. the curves for 
$E_\uparrow=-1.2,\,-2.4$). Here we find $z^2 \gtrsim 0.95$, which explains
the small deviation of the slave-boson order parameter from its
Hartree-Fock counterpart (see inset of Fig.~\ref{fig_pd}).    

\begin{figure}[t]
  \centering 
  \includegraphics[width=\linewidth,clip]{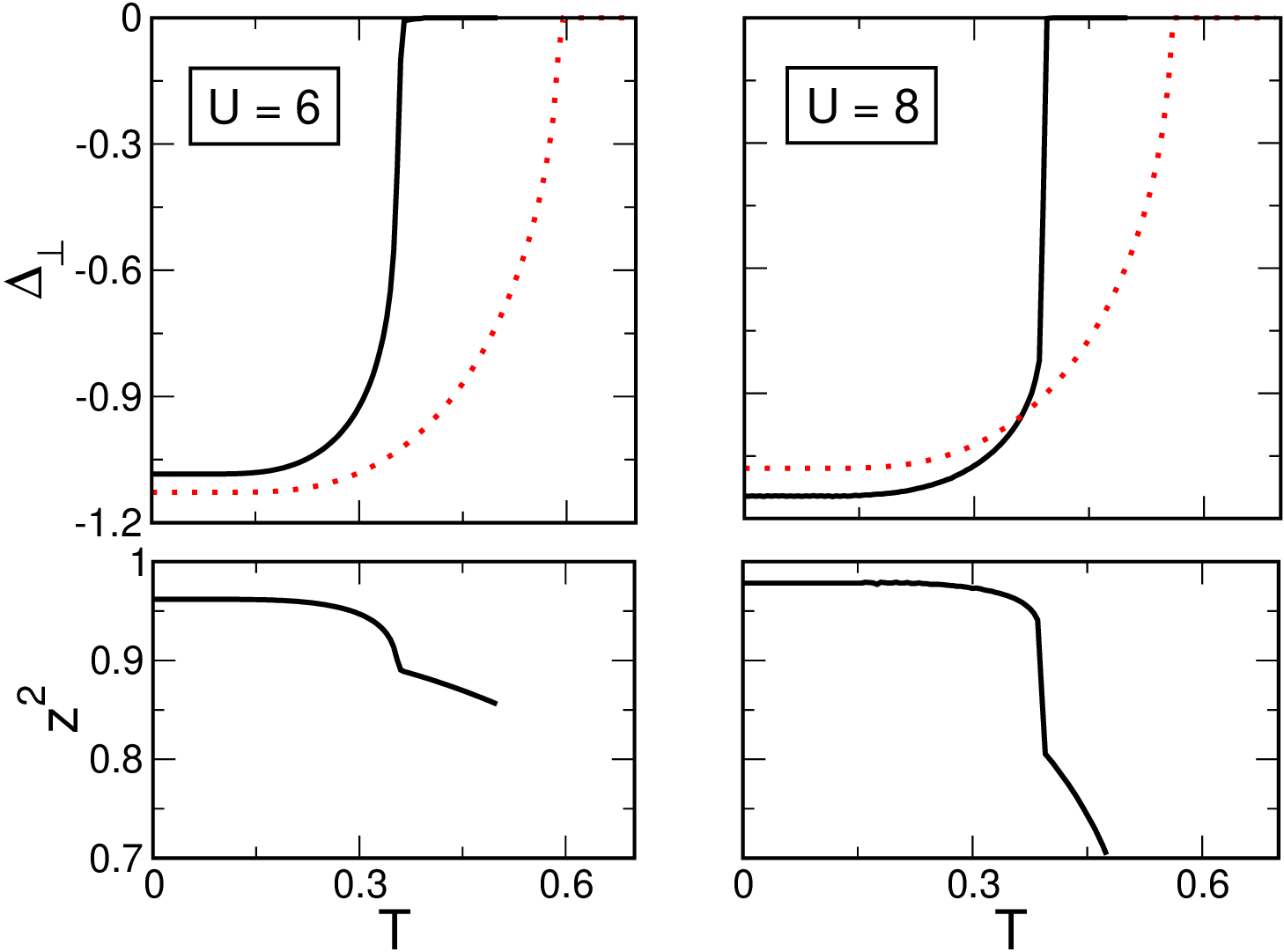}
  \caption{(Color online) $T$-dependence of the 
EI order parameter $\Delta_\perp$ and of the band renormalization 
$z^2$ at fixed Coulomb interaction $U=6$ (left-hand panels) 
and $U=8$ (right-hand panels). Red dotted lines show the corresponding 
Hartree-Fock data, where $z^2=1$. Band-structure parameters are 
the same as in Fig.~\ref{fig_pd}.\label{fig_opz2vsT}}
\end{figure}

Next we discuss the finite-temperature behavior. 
The variation of the EI order parameter and of the band-renormalization
factor with $T$ at fixed $U$ is displayed in Fig.~\ref{fig_opz2vsT}. 
Most important, in comparison with the Hartree-Fock data, 
the critical temperature for the EI-semimetal/semiconductor phase transition
is significantly reduced  (see also Fig.~\ref{fig_pd}). 
Looking at $z^2(T)$, this may be attributed to the more precise  
treatment of correlations and occupation number fluctuations.
At $T_c$, the order parameter vanishes, and we observe a cusp in
$z^2$. Enhancing, above $T_c$, the temperature further,  
the band renormalization goes on, where $z^2$ now always
decreases with increasing $U$.

\begin{figure}[t]
  \centering 
\includegraphics[width=0.48\linewidth,clip]{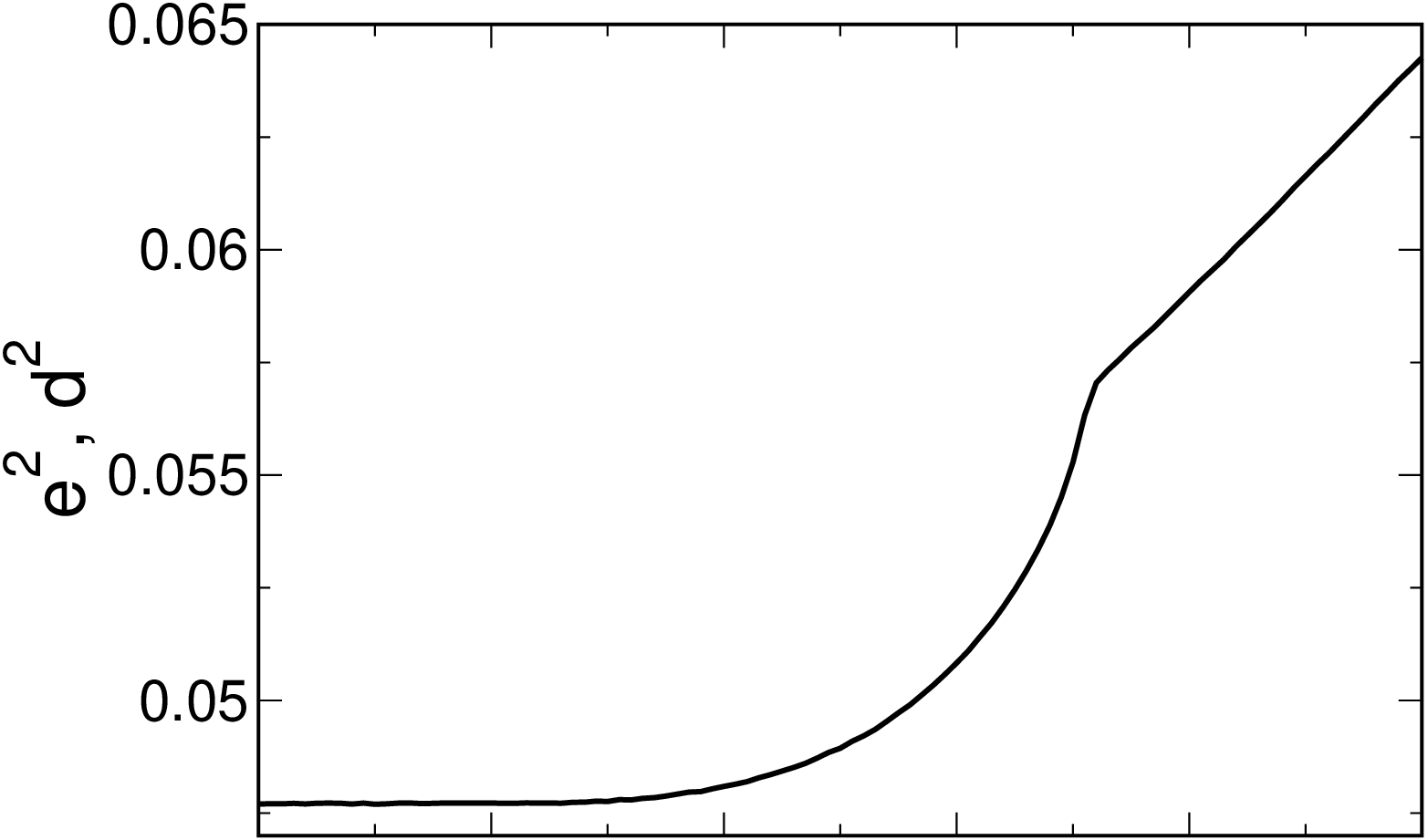}\hspace*{0.2cm}
\includegraphics[width=0.48\linewidth,clip]{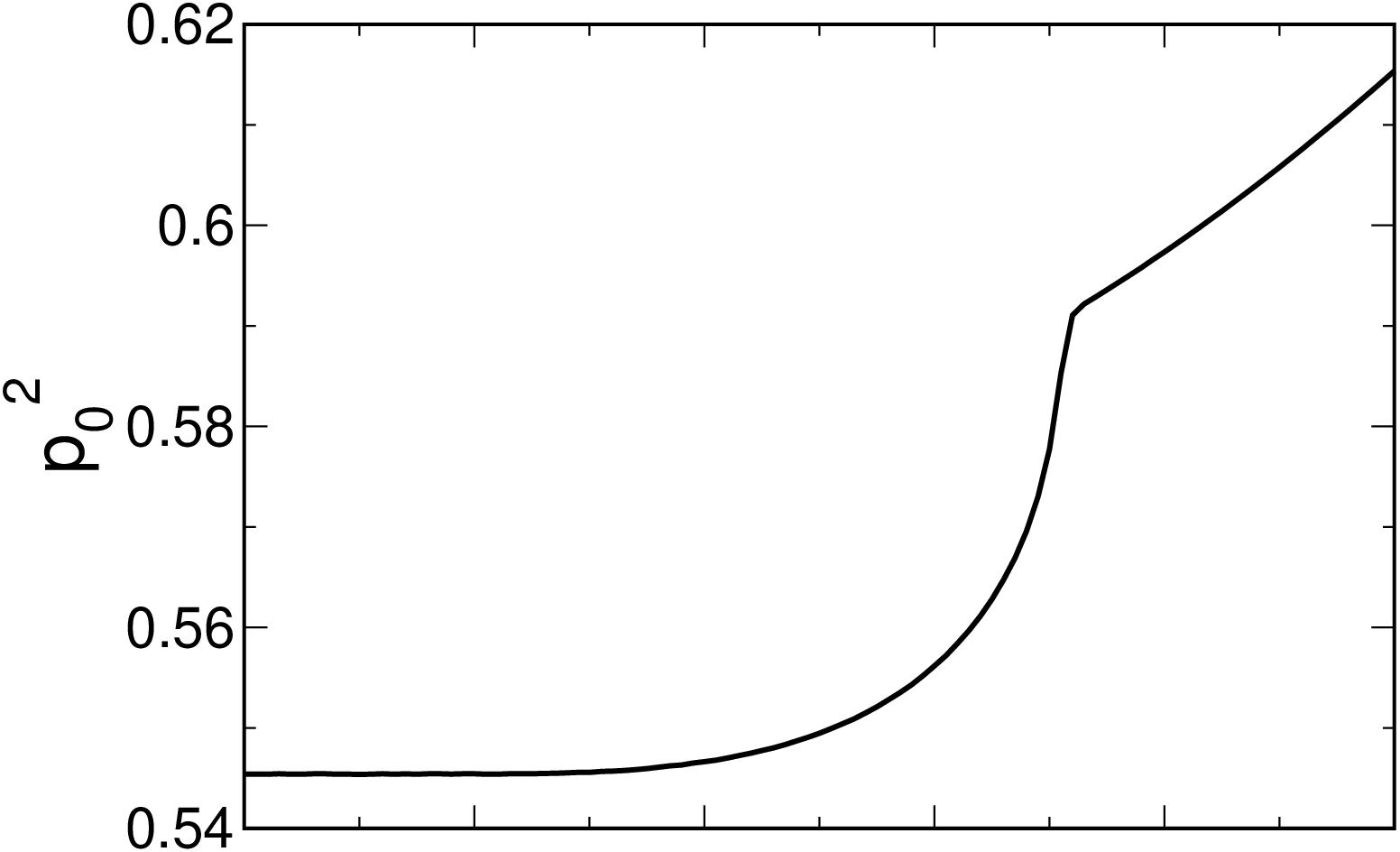}\\[0.1cm]
\includegraphics[width=0.48\linewidth,clip]{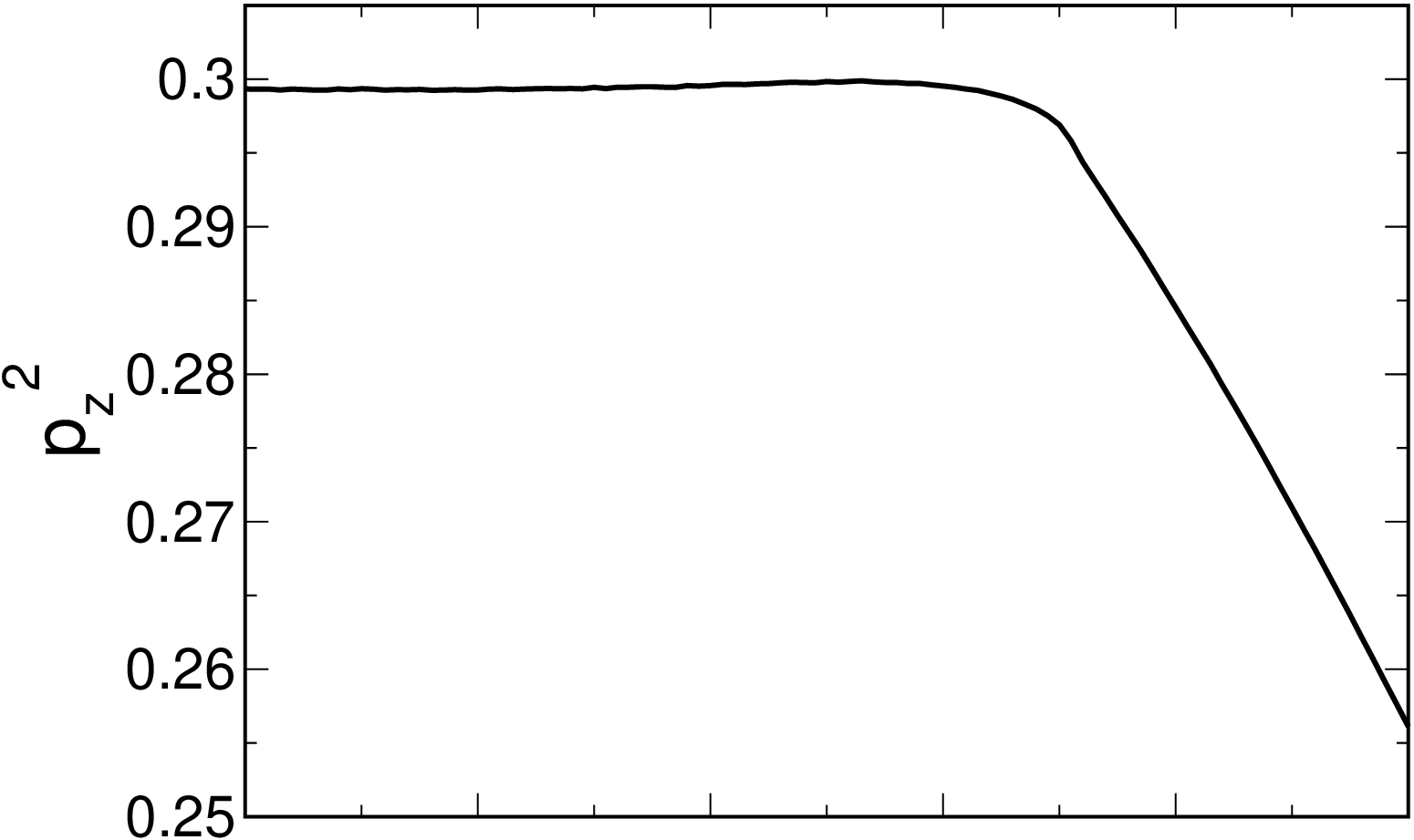}\hspace*{0.2cm}
\includegraphics[width=0.48\linewidth,clip]{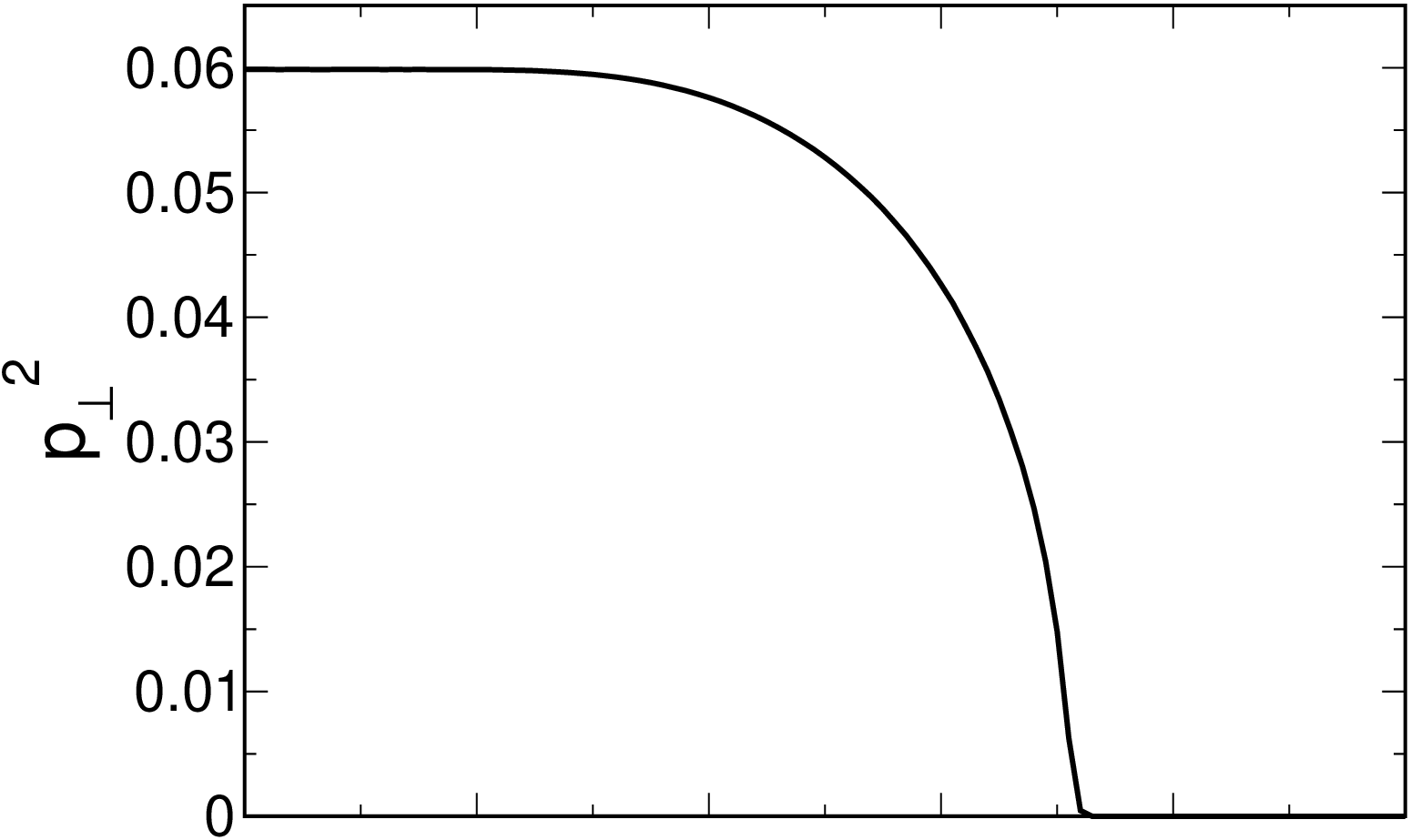}\\
\hspace*{0.2cm}\includegraphics[width=0.48\linewidth,clip]{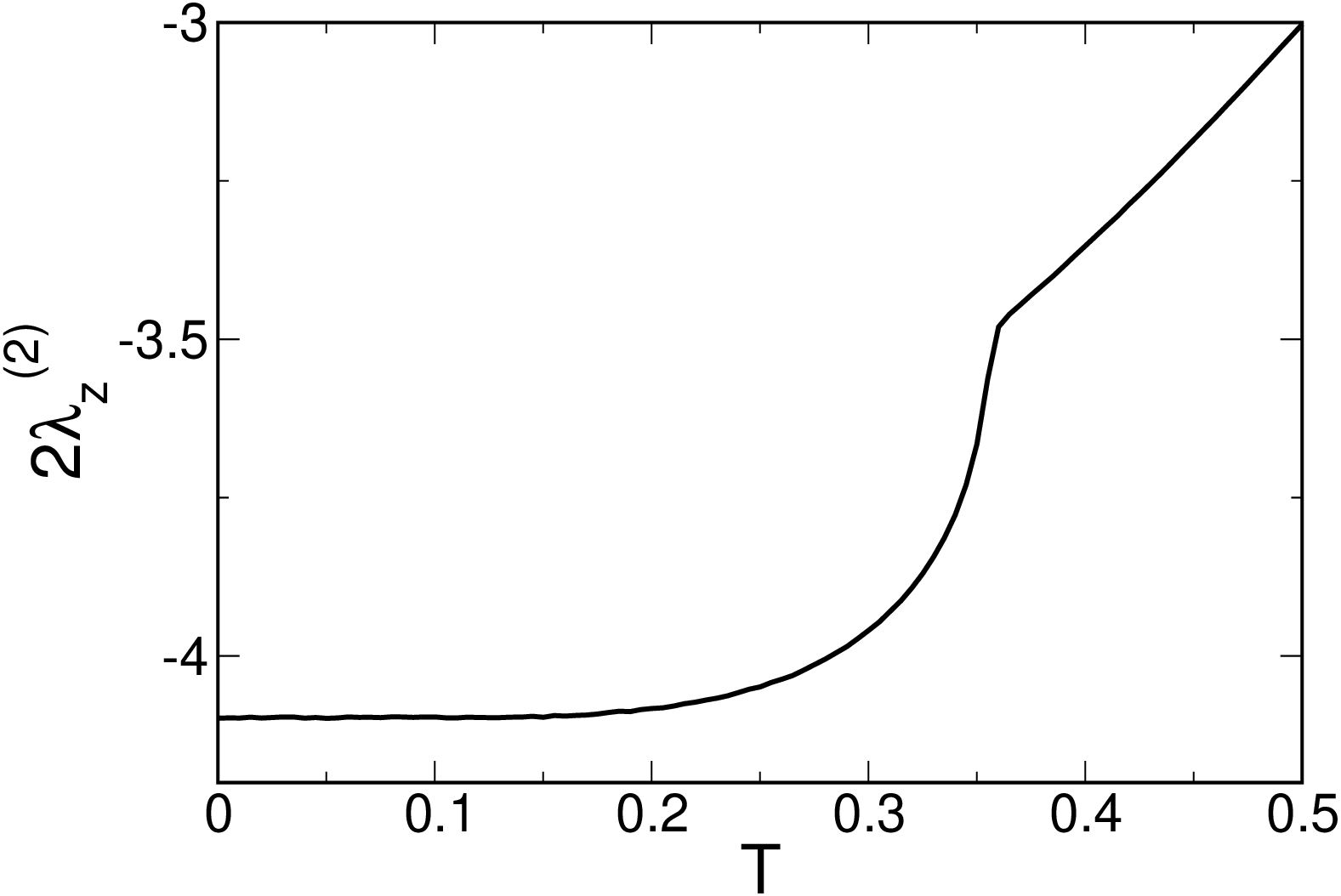}\hspace*{0.2cm}
\includegraphics[width=0.48\linewidth,clip]{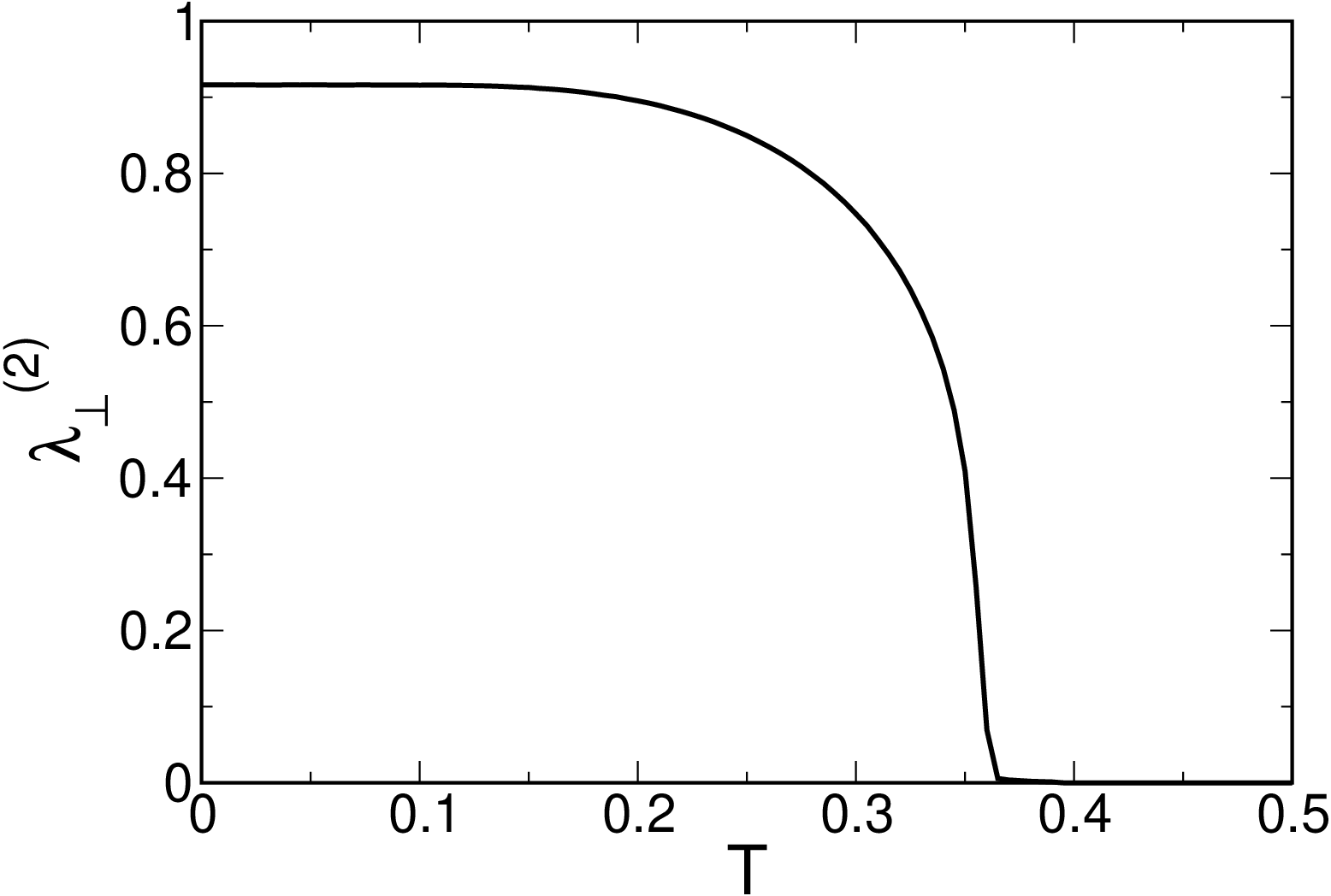}
  \caption{$T$-dependence of the 
various slave-boson fields for $U=6$, $E_\uparrow=-2.4$, and 
$\kappa =-0.8$.\label{fig_scvsT}}
\end{figure}

Figure~\ref{fig_scvsT} shows the temperature dependencies
of the various slave-boson fields and Lagrange parameters for $U=6$
(corresponding to the left-hand panel of Fig.~\ref{fig_opz2vsT}). 
As expected, $p_\perp$ and $\lambda^{(2)}_\perp$ are monotonously 
decreasing functions of $T$, with  $p_\perp(T_c)=\lambda^{(2)}_\perp(T_c)=0$.
The other fields exhibit a cusp structure at $T_c$. 
At higher temperatures the probability of finding double occupied 
sites and empty sites increases. At the same time, we find  less singly 
occupied sites ($\propto (p_0^2+p_z^2)$), which means that the increase 
of $p_0^2$ is overcompensated by the reduction of $p_z^2$, indicating 
a more balanced occupation of $f$ and $c$ sites.

\begin{figure}[t]
  \centering 
\includegraphics[width=0.48\linewidth,clip]{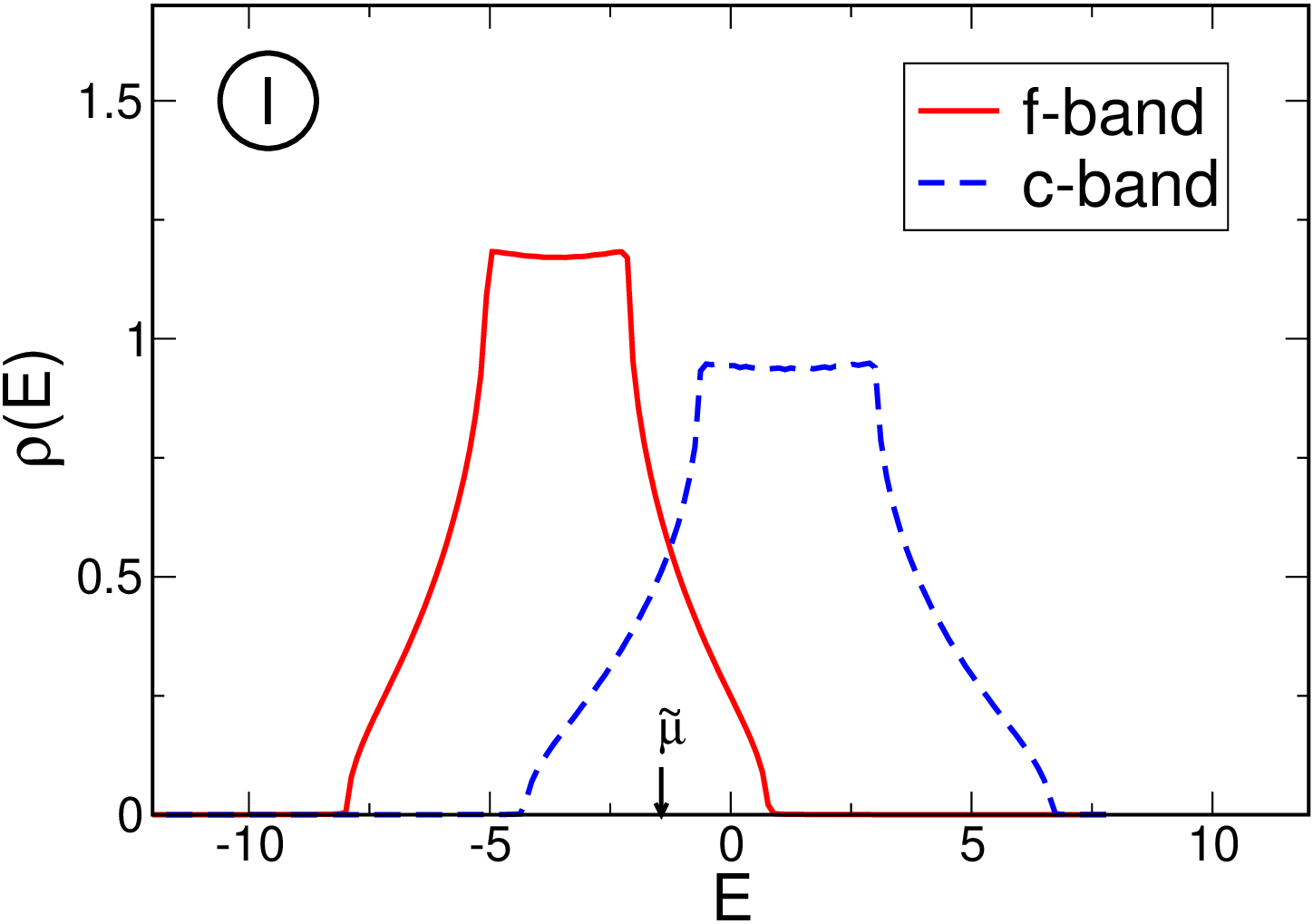}\hspace*{0.2cm}
\includegraphics[width=0.48\linewidth,clip]{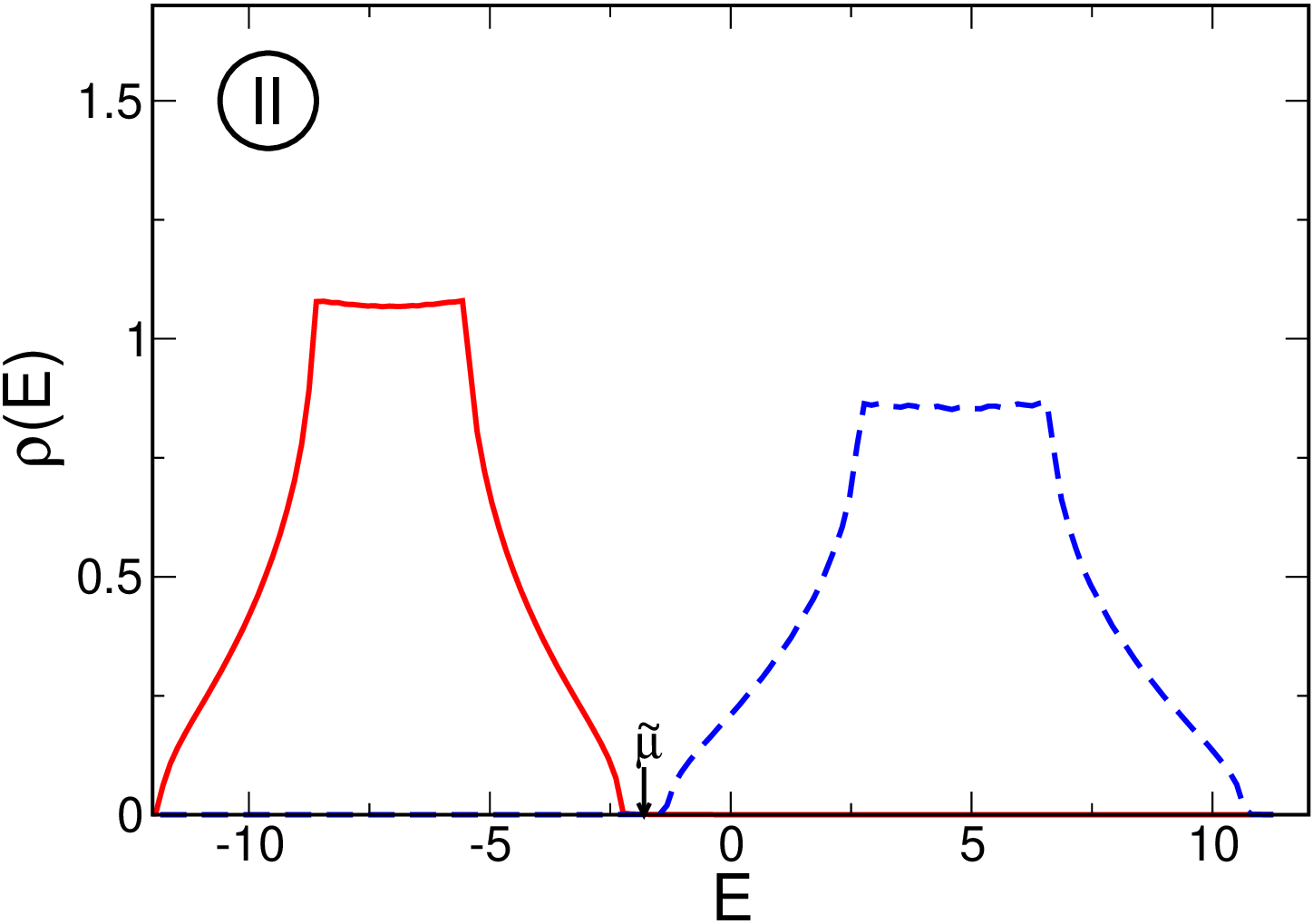}\\[0.1cm]
\includegraphics[width=0.48\linewidth,clip]{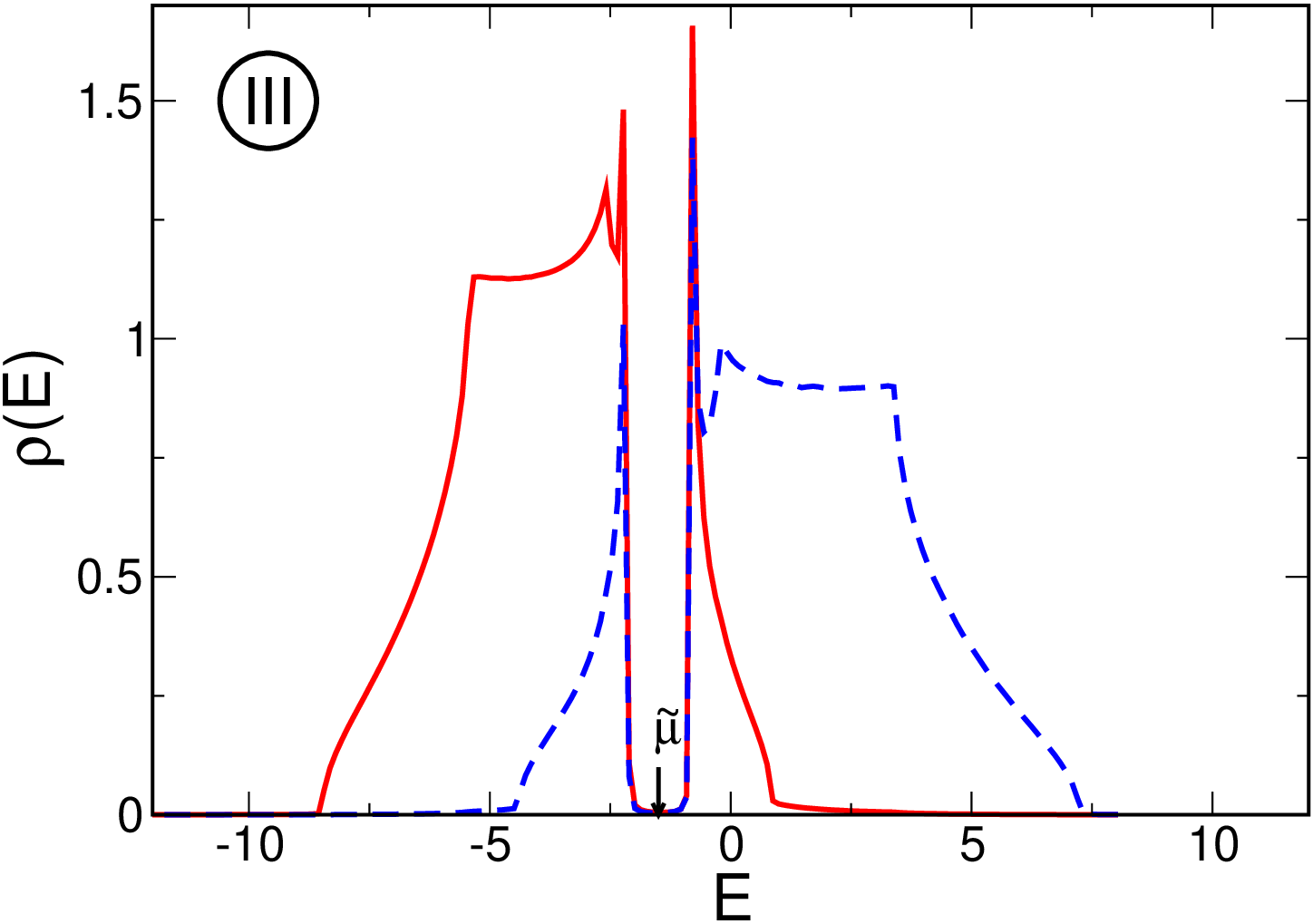}\hspace*{0.2cm}
\includegraphics[width=0.48\linewidth,clip]{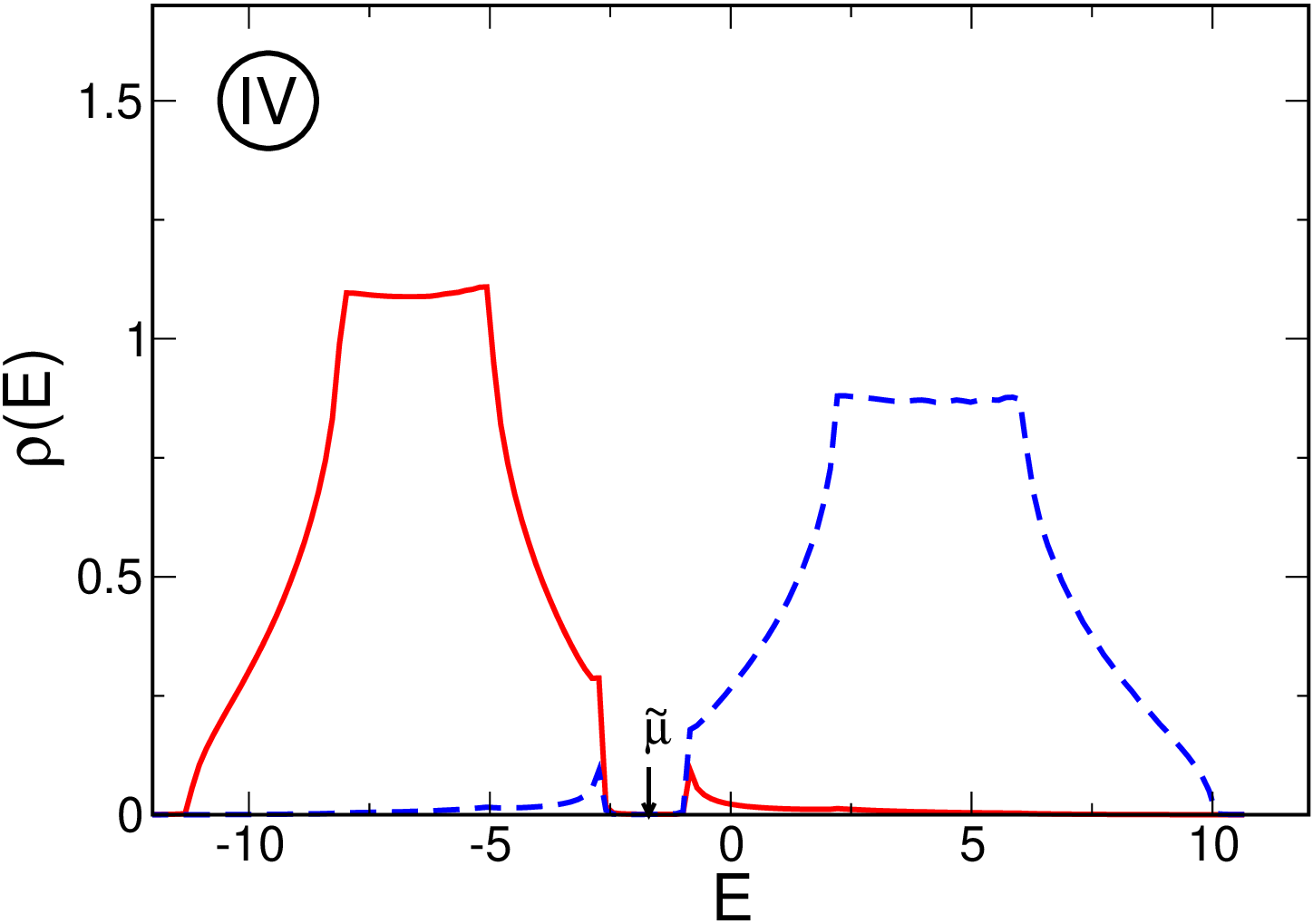}
  \caption{(Color online) Partial densities of states for $f$-band 
(red solid curves) and $c$-band (blue dashed curves) 
  electrons at the points ($U=4.8,\,T=0.45$),  
(8.5, 0.45), (4.8, 0), (8.5, 0) marked by I-IV in Fig.~\ref{fig_pd}. 
Band-structure parameters are $E_\uparrow=E_f=-2.4$ and $t_\uparrow=t_f=-0.8$.}
  \label{fig_ldos}
\end{figure}

Finally, we analyze the partial $f$  and $c$ electron density of states (DOS),
$\rho_\uparrow(E)$ and $\rho_\downarrow(E)$, defined via
\begin{align}
n_\sigma=&\frac{1}{N}\sum_{\vec{k}} n_{\vec{k}\sigma}=\int dE \rho_\sigma (E)\,,\\
n_{\vec{k}\uparrow}=&\tfrac{1}{2}(1+ m_{\vec{k}})n_{\vec{k}+}
+ \tfrac{1}{2}(1- m_{\vec{k}})n_{\vec{k}-}\,,\\
n_{\vec{k}\downarrow}=&\tfrac{1}{2}(1-m_{\vec{k}})n_{\vec{k}+}
+ \tfrac{1}{2}(1+ m_{\vec{k}})n_{\vec{k}-}\,,
\end{align}
where $n_{(\vec{k})\sigma}$ are the corresponding particle densities.
Figure~\ref{fig_ldos} gives $\rho_{\uparrow,\downarrow}(E)$
at the characteristic $U$-$T$ points marked by I-IV in the phase diagram 
of Fig.~\ref{fig_pd}. Obviously, the high-temperature phase may be
viewed as a metal/semimetal (panel I) or a small-gap semiconductor (panel II) 
in the weak-to-intermediate or strong Coulomb-attraction regime.
Accordingly, the EI phase at low temperatures shows different
characteristics as well. As can be seen from panel III, a correlation-induced 
`hybridization' gap opens in the DOS with $n_{-}=1$ ($n_{+}=0$) at $T=0$, 
indicating EI long-range order.   
The pronounced $c$--$f$ state mixing and strong enhancement of 
the DOS at the upper/lower valence/conducting band 
edges reminds a BCS-like pairing evolving from a (semi-) metallic state 
with a large Fermi surface above $T_c$. By contrast, the zero-temperature DOS 
shown in panel~IV evolves from an already gapped high-temperature phase. Here, 
preformed pairs (excitons) may exist~\cite{BF06,IPBBF08}, which undergo a BEC transition at 
$T_c$. 
\section{Summary}
In this work we studied the extended Falicov-Kimball model with respect to the formation of
an exciton condensate, which is related to the problem of electronic ferroelectricity. 
Motivated by the discrepancy concerning the existence of the excitonic insulator (EI) phase
within the Hartree-Fock and scalar slave-boson approaches, we developed an $SO(2)$-invariant slave-boson theory. The main result is that our improved slave-boson
scheme is capable of describing the EI phase in a parameter region agreeing,
at zero temperature, with Hartree-Fock (and, in 2D, constrained path 
Monte Carlo) results. 
This is in striking contrast
to recent findings by the scalar slave-boson approach~\cite{Br08}, which fails to detect
the EI phase in the case of non-degenerate $f$ and $c$ orbitals. The agreement of the
zero-temperature semimetal$\to$EI and EI$\to$band-insulator transition points with 
the Hartree-Fock and Monte Carlo values is ascribed to a rather weak
band renormalization at $T=0$. At finite temperature, band-renormalization
effects due to electronic correlations and particle number fluctuations become important, and, as a result,
our slave-boson theory yields significantly lower transition temperatures
than Hartree-Fock. From the analysis of the partial $f$, $c$, and quasiparticle 
densities of states, in the EI phase a crossover from a BCS-type condensate 
to a Bose-Einstein condensate of preformed excitons may be suggested.
The results of our investigations may form the basis of forthcoming studies,
e.g., on the effects of fluctuations around the saddle point, allowing the calculation 
of pseudo-spin and charge susceptibilities for the EFKM on an equal footing. 
\section*{Acknowledgments}
The authors thank A. Alvermann, B. Bucher, N. V. Phan, G. R\"opke, and H. Stolz for stimulating 
discussions. This work was supported by DFG through SFB 652. 

\end{document}